\documentclass[11pt]{abpaper}

\usepackage{xcolor,tikz,graphicx}
\usepackage{todonotes}
\usepackage{xspace}
\usepackage{relsize}
\usepackage[update,prepend]{epstopdf}
\usepackage{amsmath,amssymb}
\usepackage{cancel}
\usepackage{url}
\usepackage{booktabs}
\usepackage{bm}
\usepackage[font=small,labelfont=bf]{caption}
\usepackage[labelfont=md,skip=2pt]{subcaption}
\usepackage{placeins}
\usepackage{floatpag}
\usepackage{afterpage}
\usepackage{etoolbox}
\usepackage{nag}
\usepackage{bookmark}

\AtEndPreamble{%
  \usepackage{microtype}
  \usepackage{hyperref}
}

\newcommand{\img}[2][1.0]{\includegraphics[width=#1\textwidth]{#2}}

\newcommand{\eV}{\ensuremath{\text{e}\mspace{-0.8mu}\text{V}\xspace}}

\newcommand{\TeV}{\ensuremath{\text{T\eV}}\xspace}

\usepackage{units}

\newcommand{\pt}{\ensuremath{p_T}\xspace}

\newcommand{\alphaS}{\texorpdfstring{\ensuremath{\alpha_\mathrm{S}}}{alphaS}\xspace}

\newcommand{\Pythia}{\textsc{Pythia}\xspace}

\makeatletter
\DeclareRobustCommand{\kbd}[1]{{\texttt{#1}}}

\g@addto@macro\bfseries{\boldmath}
\makeatother

\usepackage{mathpazo}

\makeatletter
\DeclareOldFontCommand{\rm}{\normalfont\rmfamily}{\mathrm}
\DeclareOldFontCommand{\sf}{\normalfont\sffamily}{\mathsf}
\DeclareOldFontCommand{\tt}{\normalfont\ttfamily}{\mathtt}
\DeclareOldFontCommand{\bf}{\normalfont\bfseries}{\mathbf}
\DeclareOldFontCommand{\it}{\normalfont\itshape}{\mathit}
\DeclareOldFontCommand{\sl}{\normalfont\slshape}{\@nomath\sl}
\DeclareOldFontCommand{\sc}{\normalfont\scshape}{\@nomath\sc}
\makeatother

\newcommand{\Powheg}{\textsc{Powheg}\xspace}
\newcommand{\PowhegBox}{\textsc{Powheg-Box}\xspace}
\renewcommand{\Pythia}{\textsc{Pythia}\xspace}
\newcommand{\PythiaEight}{\textsc{Pythia}\,8\xspace}

\newcommand{\ptdef}{\text{\texttt{pTdef}}\xspace}
\newcommand{\pthard}{\text{\texttt{pThard}}\xspace}
\newcommand{\ptemt}{\text{\texttt{pTemt}}\xspace}

\newcommand{\pT}{\ensuremath{p_\mathrm{T}}\xspace}
\renewcommand{\pt}{\pT}
\newcommand{\ET}{\ensuremath{E_\mathrm{T}}\xspace}

\author{\textbf{Andy Buckley}\\[-2mm] {\rmfamily\smaller\emph{School of Physics \& Astronomy, Glasgow University, UK}}\\[1mm]
  \textbf{Debottam Bakshi Gupta}\\[-2mm] {\rmfamily\smaller\emph{Department of Physics, Louisiana Tech University, USA}}
}
\title{\setstretch{1} \Powheg--\Pythia matching scheme effects in\\ NLO simulation of dijet events}
\preprints{MCnet-16-34} 
\date{\today}

\begin{document}

\begin{abstract}
  One of the most important developments in Monte Carlo simulation of collider
  events for the LHC has been the arrival of schemes and codes for matching of
  parton showers to matrix elements calculated at next-to-leading order in the
  QCD coupling. The \Powheg scheme, and particularly its implementation in the
  \PowhegBox code, has attracted most attention due to ease of use and effective
  portability between parton shower algorithms.

  But formal accuracy to NLO does not guarantee predictivity, and the
  beyond-fixed-order corrections associated with the shower may be
  large. Further, there are open questions over which is the ``best'' variant of
  the \Powheg matching procedure to use, and how to evaluate systematic
  uncertainties due to the degrees of freedom in the scheme.

  In this paper we empirically explore the scheme variations allowed in
  \PythiaEight matching to \PowhegBox dijet events, demonstrating the effects of
  both discrete and continuous freedoms in emission vetoing details for both
  tuning to data and for estimation of systematic uncertainties from the
  matching and parton shower aspects of the \PowhegBox+\PythiaEight generator
  combination.
\end{abstract}

\section*{Introduction}

One of the most important recent developments in Monte Carlo simulation of
collider events has been the arrival of schemes and codes for consistent
parton-shower dressing of partonic hard process matrix elements calculated at
next-to-leading order in the QCD coupling.

It is now possible to simulate fully exclusive event generation in which the
parton shower (PS) is smoothly matched to matrix element (ME) calculations
significantly improved over the leading-order Born level, and the modelling
further improved by non-perturbative modelling aspects such as hadronization and
multiple partonic interactions (MPI).  Tools providing these improvements
include both ``multi-leg LO'', exemplified by the Alpgen~\cite{Mangano:2002ea},
MadGraph~\cite{Alwall:2011uj}, and Sherpa\,1~\cite{Gleisberg:2008ta} codes; the
``single-emission NLO'' codes such as \PowhegBox~\cite{Frixione:2007vw} and
(a)MC@NLO~\cite{Frixione:2002bd,Alwall:2014hca}; and the latest generation in
which both modes are combined into shower-matched multi-leg NLO: Sherpa\,2 and
MadGraph5-aMC@NLO~\cite{Frederix:2012ps}.

While the technical and bookkeeping details in these algorithms for combination
of different-multiplicity matrix elements and parton showers are formidable, and
their availability has revolutionised the approaches taken to physics analysis
during the LHC era, there remain constant questions about how to evaluate the
uncertainties in the methods. Which generator configuration choices are absolute
and unambiguous, and which have degrees of freedom which can be exploited either
for more accurate data-description (of primary interest to new physics search
analyses) or to construct a theory systematic uncertainty in comparisons of QCD
theory to data (the ``Standard Model analysis'' attitude).  Despite the
confidence-inspiring ``NLO'' label on many modern showering generators, there
are in practice many freedoms in matching matrix elements to shower generators.

The \Powheg scheme, in particular its implementation in the \PowhegBox
code~\cite{Frixione:2007vw}, has attracted most attention due to its ease of use and
formal lack of dependence on the details of the parton shower used. But formal
accuracy to NLO does not guarantee predictivity, and the beyond-fixed-order
corrections induced by the shower procedure may be large. Further, questions
remain over which is the ``best'' variant of the \Powheg matching procedure to
use, and how to evaluate systematic uncertainties due to the degrees of freedom
in the scheme. Since the extra parton production in \Powheg real-emission events
suppresses the phase space for parton shower emission, use of such ME--PS
matching can lead to underestimation of total systematic uncertainties unless
the matching itself is considered as a potential source of uncertainty in
addition to the \Powheg matrix element scales and the (suppressed) \Pythia
parton showers.

In this paper we empirically explore the scheme variations allowed in
\PythiaEight~\cite{Sjostrand:2014zea} matching to \PowhegBox dijet~\cite{Alioli:2010xa}
events, demonstrating the potentially disastrous effects of ``reasonable''
matching choices and the remaining tuning freedom for optimal data description.

\section{\Powheg matching variations in \PythiaEight}

The original and simplest approach to showering \Powheg events is to start
parton shower evolution at the characteristic scale declared by the input
event. Since the \Powheg formalism works via a shower-like Sudakov form factor,
and both \PowhegBox and the \Pythia parton showers produce emissions ordered in
relative transverse momentum, this approach seems intuitively correct. But in
fact the definition of ``relative transverse momentum'' is not quite the same
between the two codes and hence this approach may double-count some phase-space
regions, and fail to cover others entirely.

\Pythia's answer to this is to provide machinery for ``shower vetoing'', i.e. to
propose parton shower emissions over all permitted phase space (including above
the input event scale, up to the beam energy threshold) according to \Pythia's
emission-hardness definition, but to veto any proposal whose \Powheg definition
of ``hardness'' is above the threshold declared by or calculated from the input
event. This machinery is most easily accessed via the \kbd{main31} \PythiaEight
example program.

But there are still ambiguities, since the \Powheg method itself does not
prescribe exactly what form of vetoing variable should be used. \PythiaEight
makes three discrete choices available for calculation of the \Powheg-hardness
scales to be used in the shower-emission vetoing.  These are controlled via the
configuration flags \ptdef, \ptemt, and \pthard, which respectively determine
the variable(s) to be used to define ``hardness'' in the \Powheg shower-veto
procedure, the partons to be used in computing their values, and the cut
value(s) to be used in applying that hardness veto to the calculated hardness
variables.

The available variations of these scale calculations range from comparison of
the matched emission against only a limited subset of event particles at one
extreme, to comparison with all available particles at the other. This typically
produces a spectrum of scale values to characterise a shower emission, from
maximal scales at one end to minimal scales at the other, with a spectrum of
in-between values from hybrid approaches. The effect on shower emission vetoing
depends on the combination of the scale calculated for each proposed shower
emission and the scale threshold determined from the input event.

Testing all options of all three scales simultaneously would produce 50 or so
predictions to be compared and disambiguated: not a pleasant task for either us
or the reader! So we instead take a divide-and-conquer approach, first focusing on
the \ptemt scale alone since it has been observed to produce large effects in
many observables. We then proceed via a reduced set of preferred \ptemt schemes,
on which to study further variations.

In all the following comparisons and discussion the combination of scale
calculation schemes are represented by an integer tuple
$HED = (\pthard, \ptemt, \ptdef)$. The default \PythiaEight \Powheg matching
configuration is $HED = 201$ in this notation.

\subsection{Methodology}

All the plots shown in this study were computed using ATLAS and CMS jet analyses
encoded in the Rivet~2.4~\cite{Buckley:2010ar} analysis system. All such
available analyses at the time of the study used $pp$ events with
$\sqrt{s} = 7~\TeV$.  10~million input events for the analyses were generated in
LHE format~\cite{Alwall:2006yp} by \PowhegBox~v2\,r3144 and processed in
parallel through \Pythia\,8.212 using the \kbd{main31} example program, default
tune \& PDF, and HepMC event record output~\cite{Dobbs:2001ck}.

The full set of Rivet analyses used is listed in Table~\ref{tab:analyses},
giving a comprehensive view of the effects of matching scheme variations across
the public LHC measurements of hadronic jet production. For obvious reasons of
space and exposition, in this paper we only show a small representative subset
of these plots, but all 920 (!) have been considered in the discussion of
observed effects.

\begin{table}[tbp]
  \centering
  \begin{tabular}{lll}
    \toprule
    Rivet analysis name & Description \& citation \\
    \midrule

    \textit{ATLAS analyses} & \\

    \kbd{ATLAS\_2014\_I1326641} & 3-jet cross-section~\cite{Aad:2014rma} \\

    \kbd{ATLAS\_2014\_I1325553} & Inclusive jet cross-section~\cite{Aad:2014vwa} \\

    \kbd{ATLAS\_2014\_I1307243} & Jet vetoes and azimuthal decorrelations in dijet events~\cite{Aad:2014pua} \\

    \kbd{ATLAS\_2014\_I1268975} & High-mass dijet cross-section~\cite{Aad:2013tea} \\

    \kbd{ATLAS\_2012\_I1183818} & Pseudorapidity dependence of total transverse energy~\cite{Aad:2012dr} \\

    \kbd{ATLAS\_2012\_I1119557} & Jet shapes and jet masses~\cite{Aad:2012meb} \\

    \kbd{ATLAS\_2012\_I1082936} & Inclusive jet and dijet cross-sections~\cite{Aad:2011fc} \\

    \kbd{ATLAS\_2011\_S9128077} & Multi-jet cross-sections~\cite{Collaboration:2011tq} \\

    \kbd{ATLAS\_2011\_S9126244} & Dijet production with central jet veto~\cite{Aad:2011jz} \\

    \kbd{ATLAS\_2011\_S8971293} & Dijet azimuthal decorrelations~\cite{Aad:2011ni} \\

    \kbd{ATLAS\_2011\_S8924791} & Jet shapes~\cite{Aad:2011kq} \\

    \kbd{ATLAS\_2010\_S8817804} & Inclusive jet cross-section + dijet mass and $\chi$ spectra~\cite{Aad:2010wv} \\

    \addlinespace

    \textit{CMS analyses} & \\

    \kbd{CMS\_2014\_I1298810} & Ratios of jet \pt spectra~\cite{Chatrchyan:2014gia} \\

    \kbd{CMS\_2013\_I1224539\_DIJET} & Jet mass measurement in dijet events~\cite{Chatrchyan:2013rla} \\

    \kbd{CMS\_2013\_I1208923} & Jet \pt and dijet mass~\cite{Chatrchyan:2012bja} \\

    \kbd{CMS\_2012\_I1184941} & Inclusive dijet production as a function of $\xi$~\cite{Chatrchyan:2012vc} \\

    \kbd{CMS\_2012\_I1090423} & Dijet angular distributions~\cite{Chatrchyan:2012bf} & \\

    \kbd{CMS\_2012\_I1087342} & Forward and forward + central jets~\cite{Chatrchyan:2012gwa} \\

    \kbd{CMS\_2011\_S9215166} & Forward energy flow in dijet events~\cite{Chatrchyan:2011wm} \\

    \kbd{CMS\_2011\_S9088458} & Ratio of 3-jet over 2-jet cross-sections~\cite{Chatrchyan:2011wn} \\

    \kbd{CMS\_2011\_S9086218} & Inclusive jet cross-section~\cite{Chatrchyan:2011me} \\

    \kbd{CMS\_2011\_S8968497} & Dijet angular distributions~\cite{Khachatryan:2011as} \\

    \kbd{CMS\_2011\_S8950903} & Dijet azimuthal decorrelations~\cite{Khachatryan:2011zj} \\

    \bottomrule
  \end{tabular}
  \caption{List of Rivet analyses used for the jet observables studied in this paper.
    All analyses were performed on $pp$ data at $\sqrt{s} = 7~\TeV$.}
  \label{tab:analyses}
\end{table}

\subsection{Variation of \ptemt}

\begin{figure}[tp]
  \centering
  \subcaptionbox{}{\img[0.48]{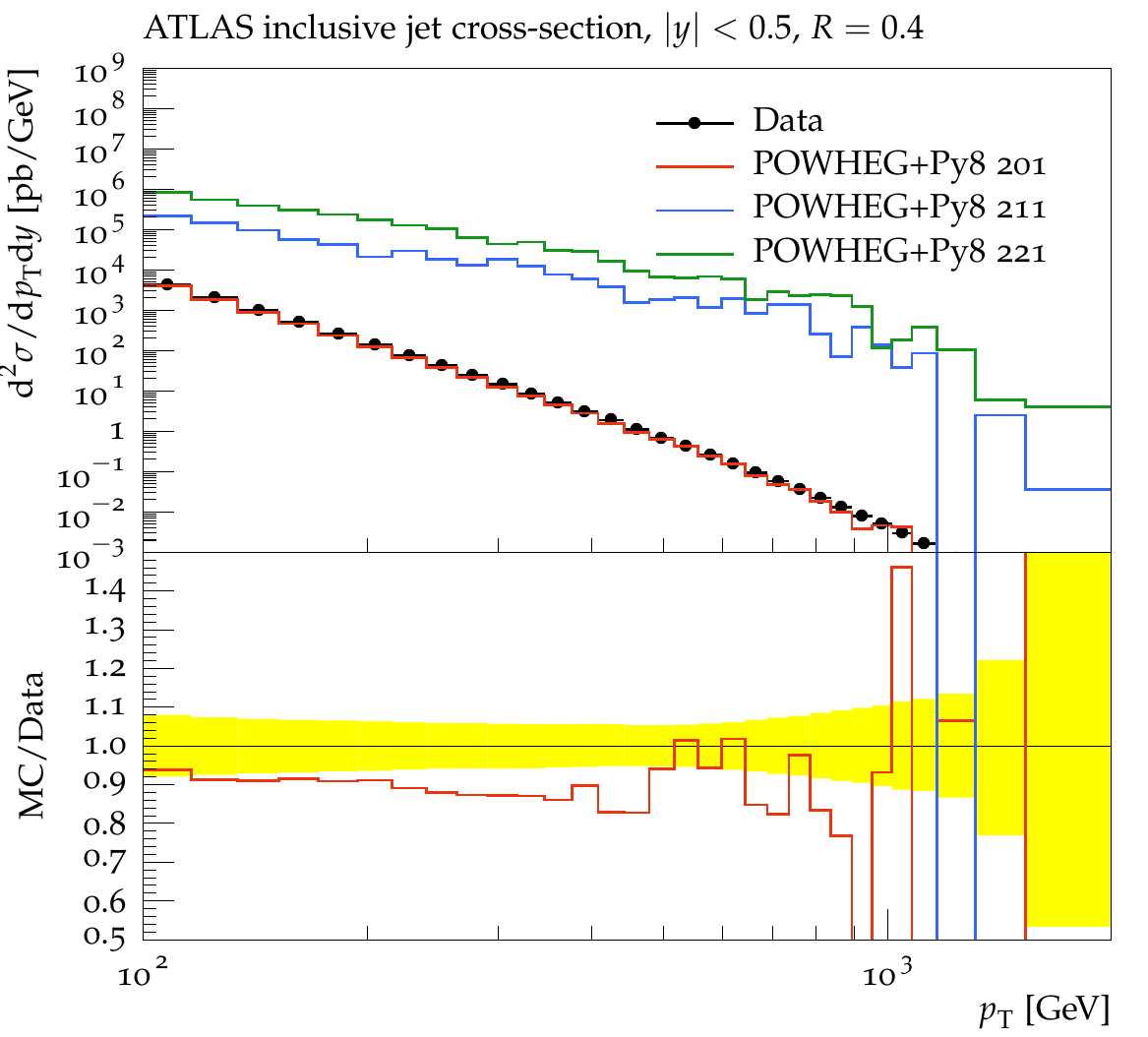}}\hfill
  \subcaptionbox{}{\img[0.48]{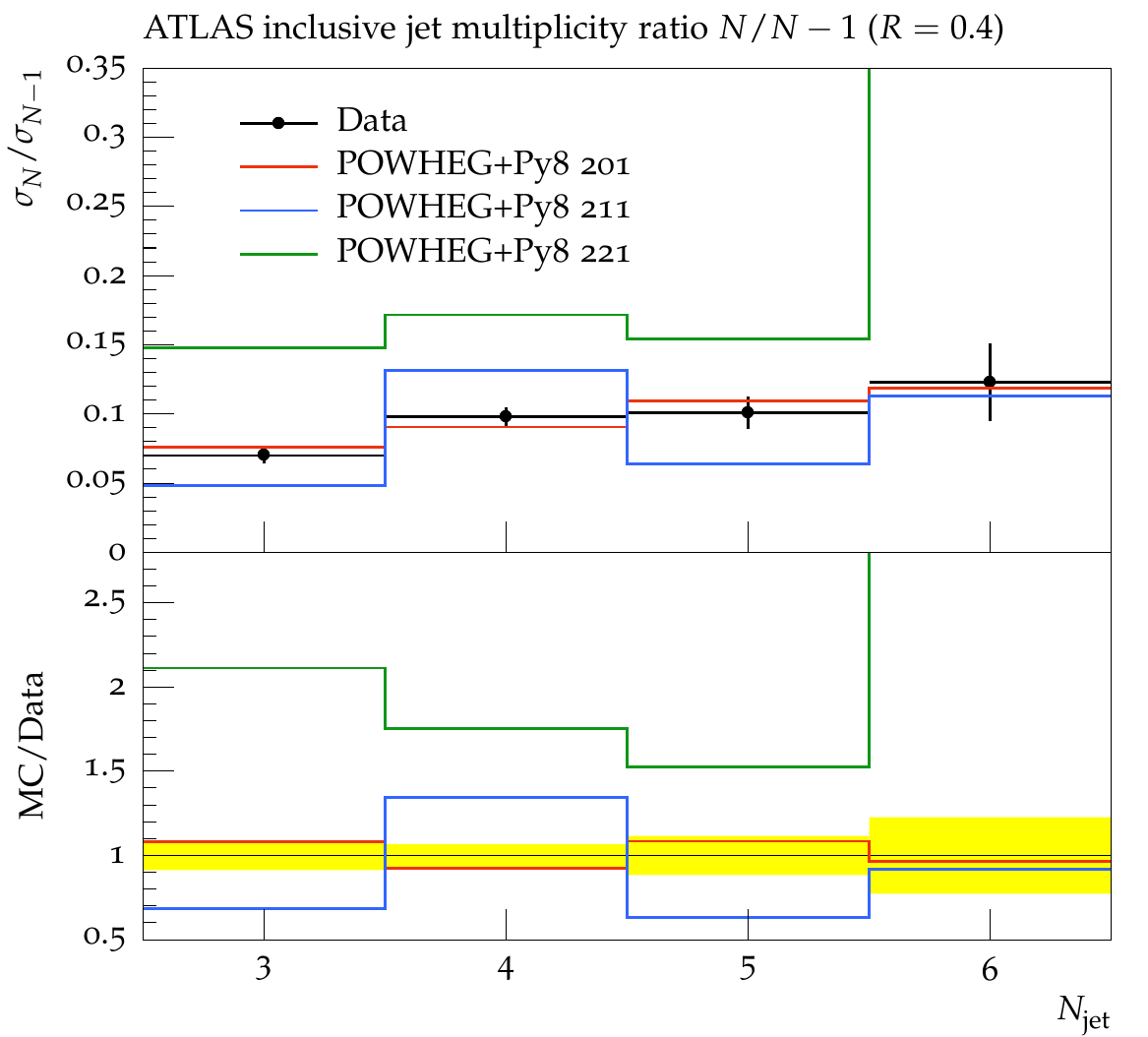}}
  \caption{Observables showing the effect of \ptemt variation. The 3-digit tuple
    used in the plot legends represents the \PythiaEight $HED$ flag combination as
    described in the text.}
  \label{fig:ptemtcmps}
\end{figure}

In Figure~\ref{fig:ptemtcmps} we show the effects of varying the \ptemt scale
calculation; for clarity only the combinations with the \kbd{main31} default
$\ptdef = 2$ and $\pthard = 1$ settings are shown, since the \ptemt effects by
far dwarf those from the two remaining degrees of freedom.

\noindent The values 0--2 of \ptemt have the following meanings, relating to the
``hardness'' of a proposed shower emission, to be compared to the vetoing
hardness cut specified by the \Powheg method and the hard-process event
kinematics:
\begin{description}
\item[0:] hardness calculated for the emitted parton only, with respect to the
  radiating parton only, and recoil effects are neglected;
\item[1:] hardness calculated for the emitted parton, computed with respect to
  \emph{all} partons (initial and final), and the minimum such value is used;
\item[2:] hardness calculated for all final-state partons, computed with respect
  to all other partons, and the minimum value used.
\end{description}
Scheme 0 is the hardness definition used by the \PowhegBox itself, in the hard
process events supplied to \Pythia, and is hence \textit{a priori} expected to
give the best matching. Since the other schemes consider hardness relative to
partons other than the emitter and take the smallest, they will in general
produce lower hardness scales for proposed shower emissions and hence fewer such
emissions will be vetoed.

The left-hand plot shows the \pt distribution of inclusive jets in the most
central rapidity bin as measured by ATLAS in 7~\TeV $pp$
collisions~\cite{Aad:2014vwa}, where the effect of using a non-default
\ptemt scheme is a cross-section overestimation by factors of 50--300. The
\Powheg matching details can hence be exceedingly important, potentially more-so
than standard systematic variations on matrix element scales and PDFs. The
right-hand plot shows the effect of \ptemt on jet multiplicity ratios: a much
smaller effect, but still a source of significant mismodelling.

It is clear that these are huge effects, utterly incompatible with the data.
The only viable combinations of $HED$ parameters have $\ptemt = 0$, i.e. the
largest of the possible values for the \Powheg hardness scales since
$\ptemt = 1$ or~2 can only be less than or equal to the $\ptemt = 0$ value. The
lower hardness values computed for \Pythia shower emissions would result in less
emission-vetoing and hence harder distribution tails.

In fact, details like jet vetos, jet shapes, and \ET flow \emph{can} be
moderately well described by the configurations which produce such large jet
mass and \pT tails. This makes sense for single-jet quantities which are at
first-order independent of overall event activity, but is less obvious for the
global event variables. It is clear, though, that from the available options the
$\ptemt = 0$ configuration is the only viable choice for dijet simulation --
while noting that it still displays a systematic data/MC discrepancy of up to
20\%.

The extent to which these different scale calculation details can affect
observables, independent of the choice of \ptdef and \pthard schemes, is
potentially disturbing. The default $\ptemt = 0$ mode gives by far the closest
agreement with data and is the definition most closely related to the NLO
subtraction used in \Powheg, but since there is no unambiguously correct
calculation scheme there must be residual uncertainty over how large the effects
of much more subtle variations in scale calculation could be. Given the obvious
sensitivity of observables to this matching scheme detail, it will be
interesting to explore whether \emph{minor} refinements to the \Pythia scale
calculation can produce more reasonable variations, particularly one which might
correct for the systematic undershooting of multi-jet mass measurements.




\subsection{Variation of \pthard and \ptdef}


\begin{figure}[tp]
  \centering
  \subcaptionbox{}{\img[0.48]{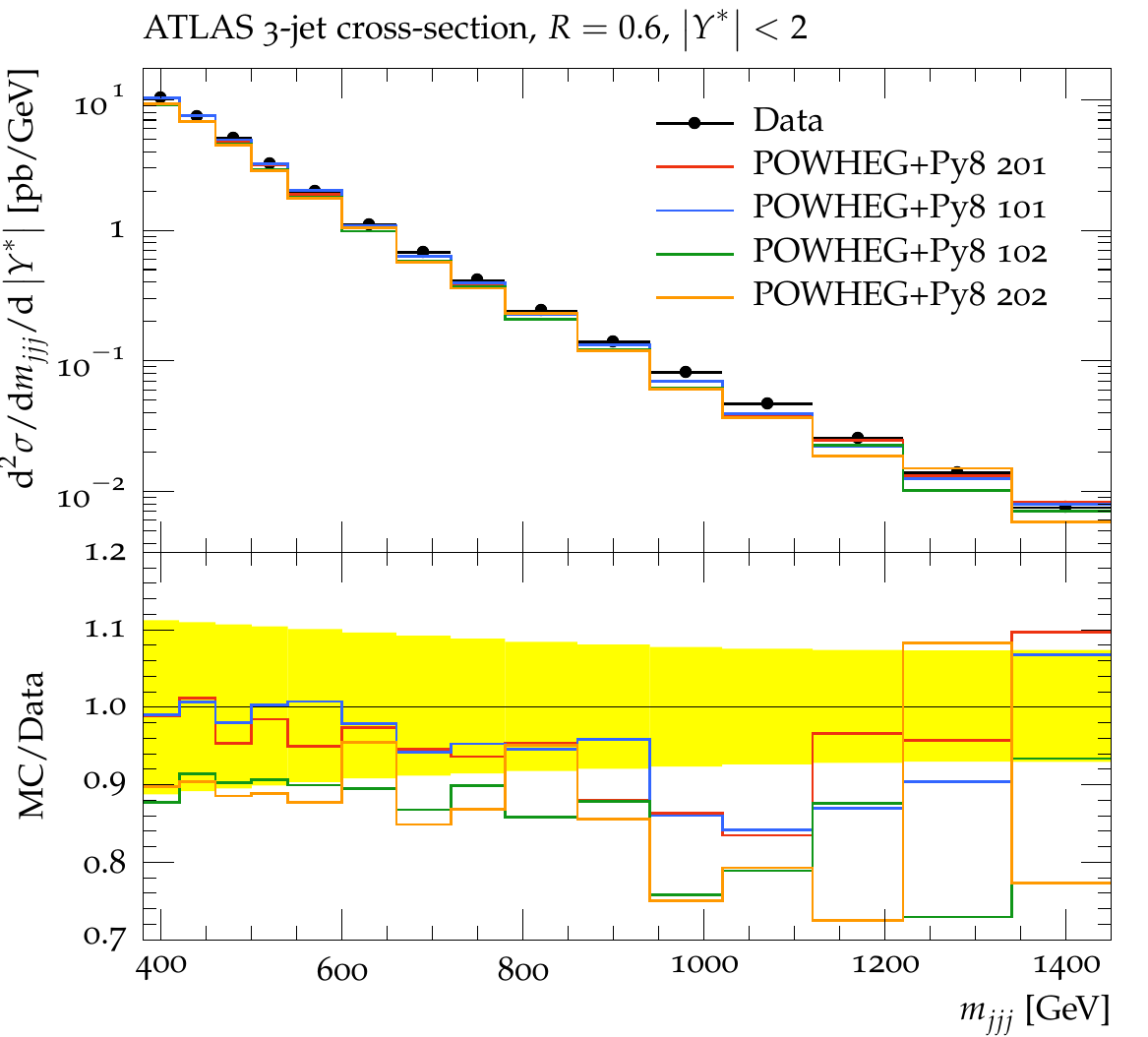}}\hfill
  \subcaptionbox{}{\img[0.48]{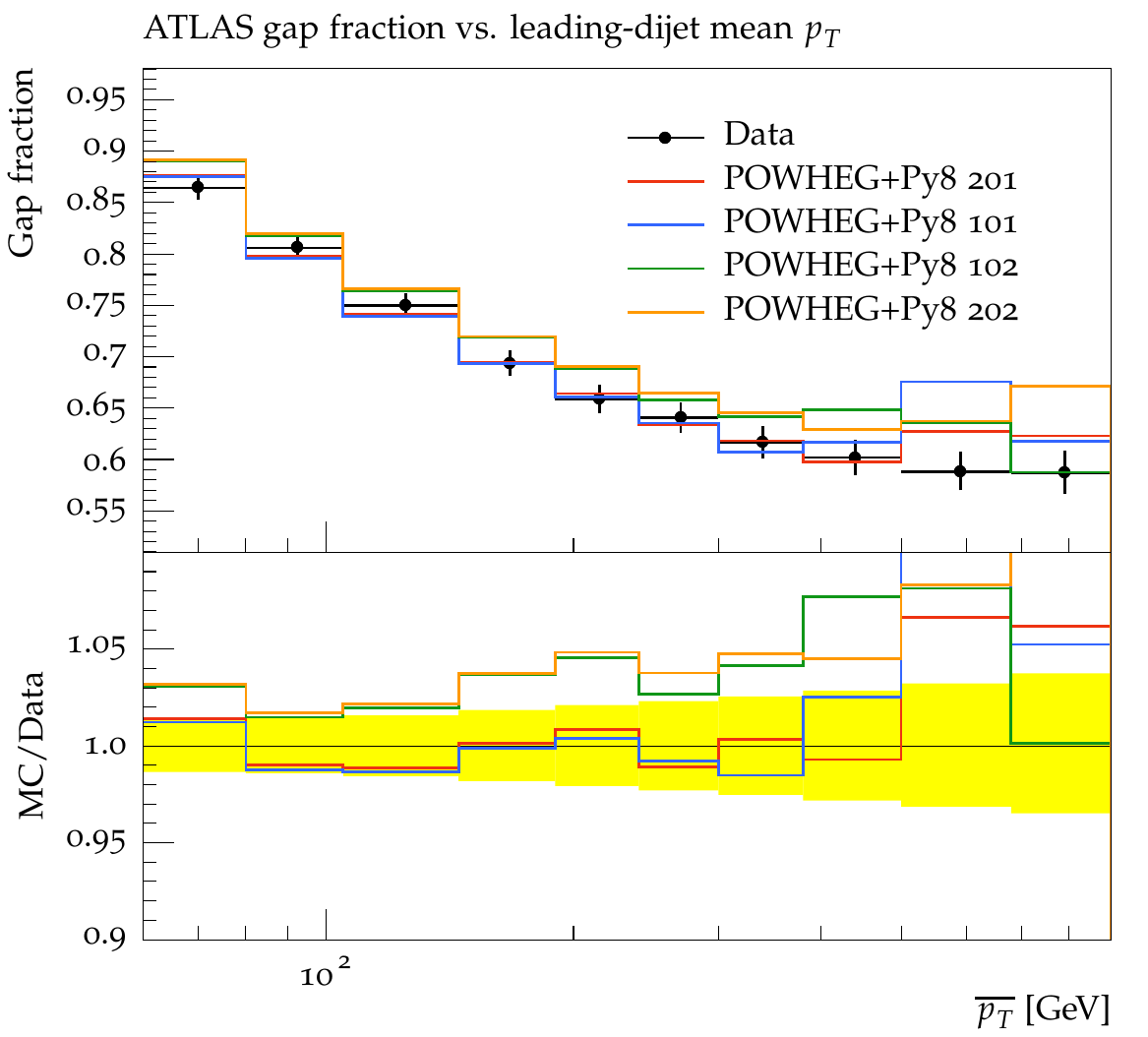}}\\[3ex]
  \subcaptionbox{}{\img[0.48]{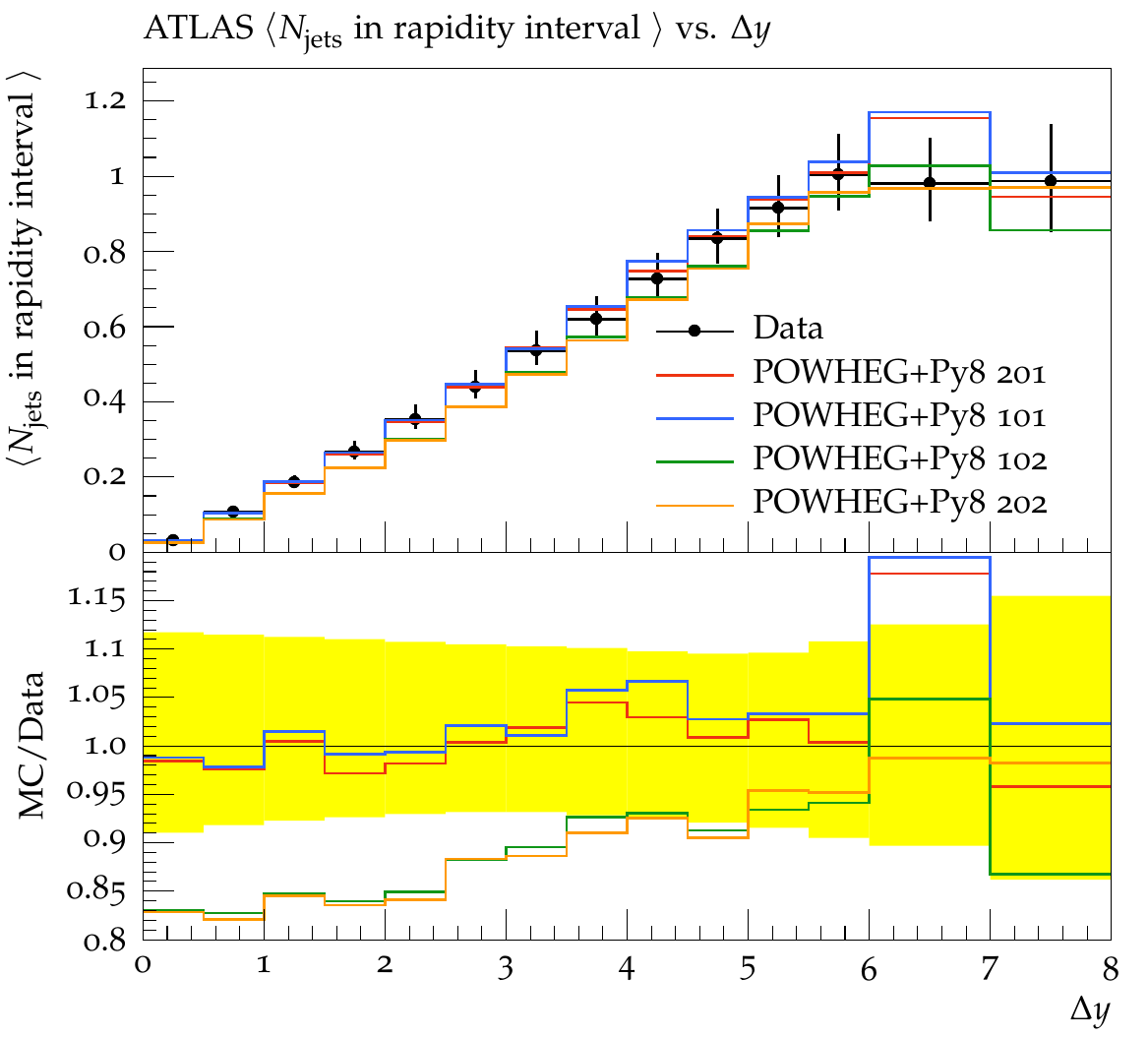}}\hfill
  \subcaptionbox{}{\img[0.48]{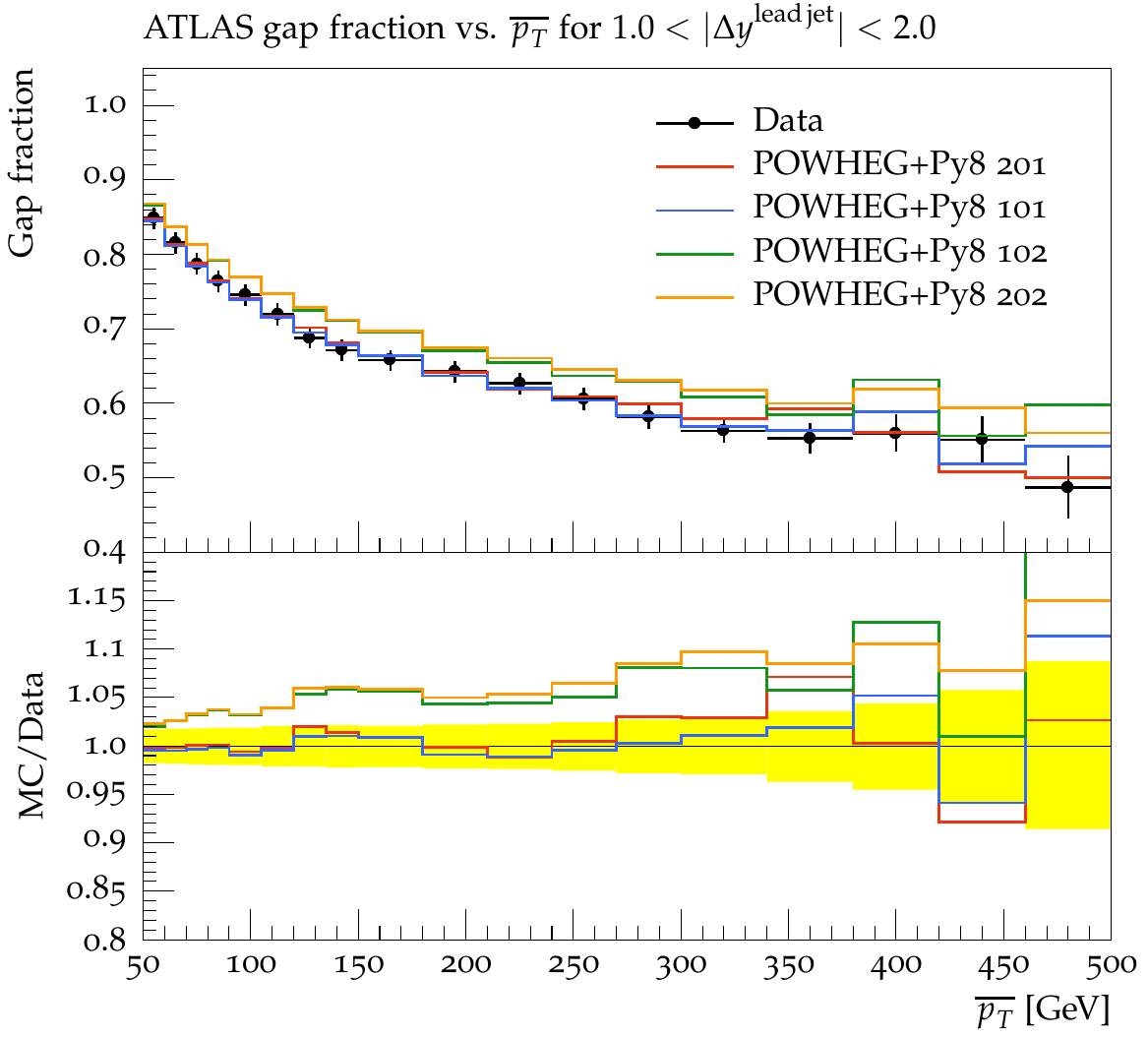}}
  \caption{Observables showing the effects of \pthard and \ptdef variations, with
    $\ptemt = 0$. The 3-digit tuple used in the plot legends represents the
    \PythiaEight $HED$ flag combination as described in the text.}
  \label{fig:ptharddefcmps1}
\end{figure}

\afterpage{\clearpage}

Having established that only the $\ptemt = 0$ configuration is viable, we now
fix its value and explore four combinations of the other two matching flags,
$\pthard \in \{1,2\}$ and $\ptdef \in \{1,2\}$; explicitly, the $HED$ tuples 201
(the \PythiaEight \texttt{main31} example program setting), 101, 202, and
102. The \pthard flag distinguishes between two methods for recalculation of the
\Powheg veto scale (as opposed to using the \kbd{SCALUP} value specified in the
LHE file): a value of 1 considers the \pt of the \Powheg emitted parton relative
to all other partons, while a value of 2 uses the minimal \pt of all final state
partons relative to all other partons. The \ptdef flag switches between using
the \PowhegBox or \PythiaEight definitions of ``\pt'' in the veto scale
calculation.

\begin{figure}[tp]
  \centering
  \subcaptionbox{}{\img[0.48]{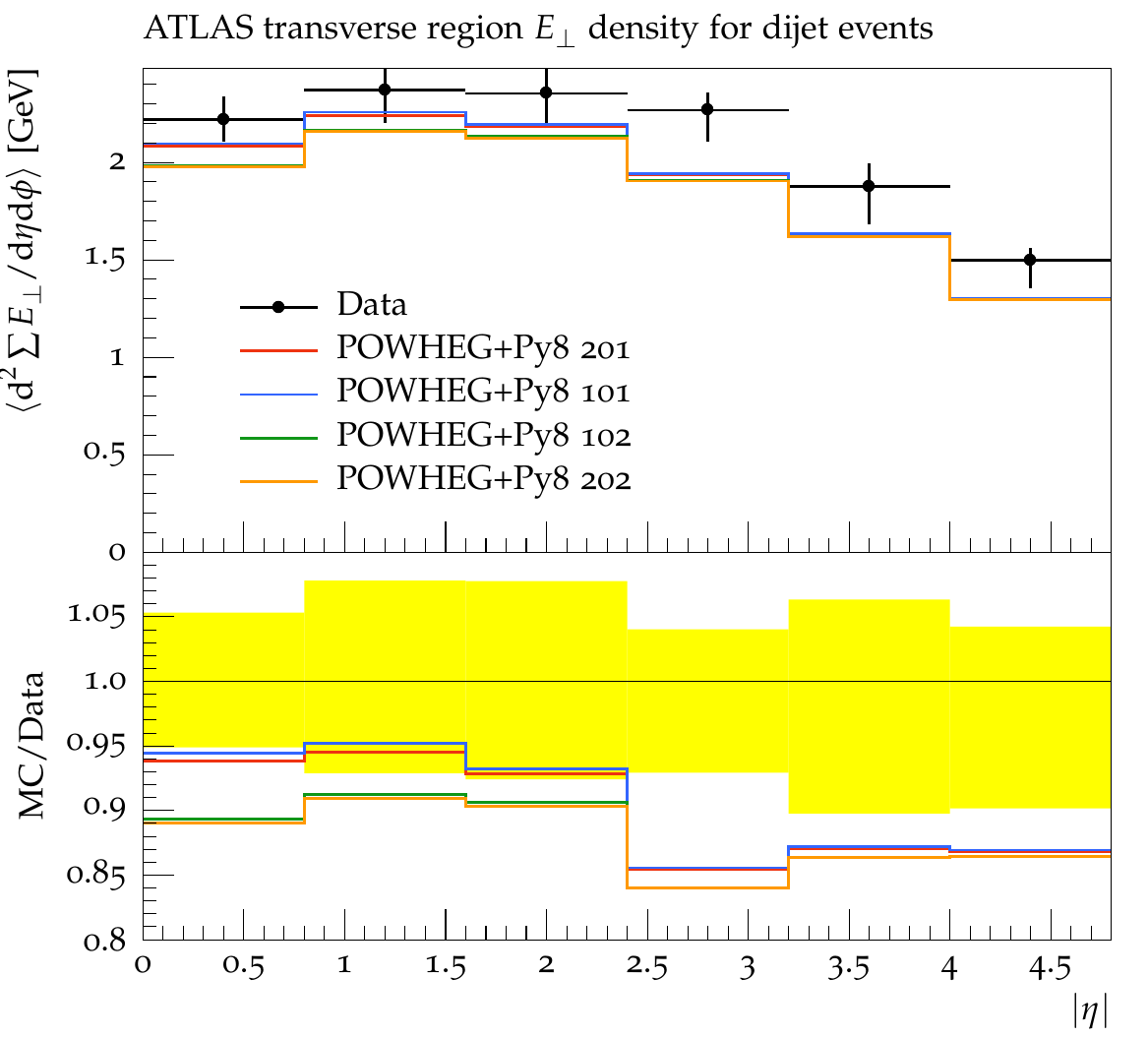}}\hfill
  \subcaptionbox{}{\img[0.48]{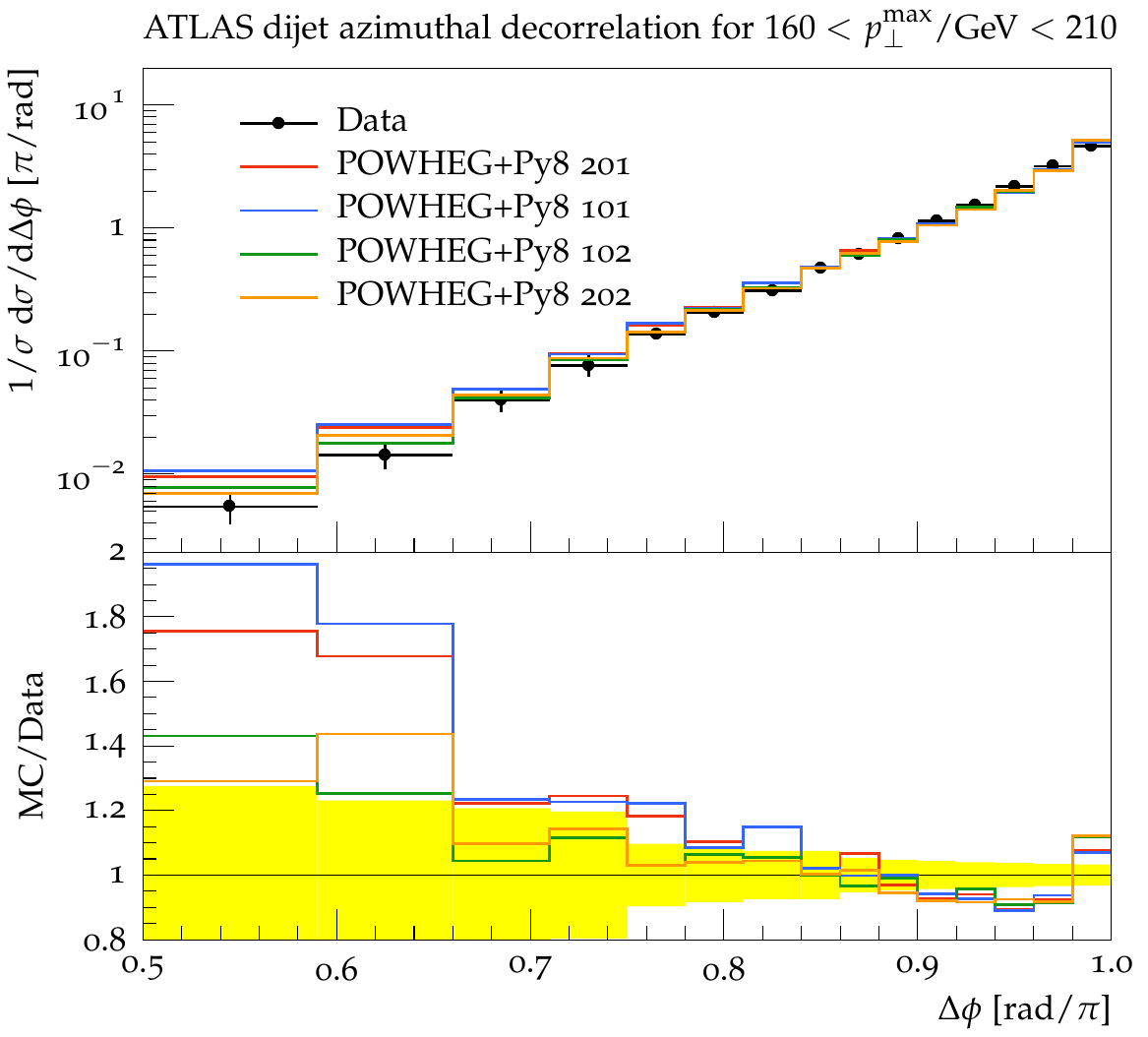}}\\[3ex]
  \subcaptionbox{}{\img[0.48]{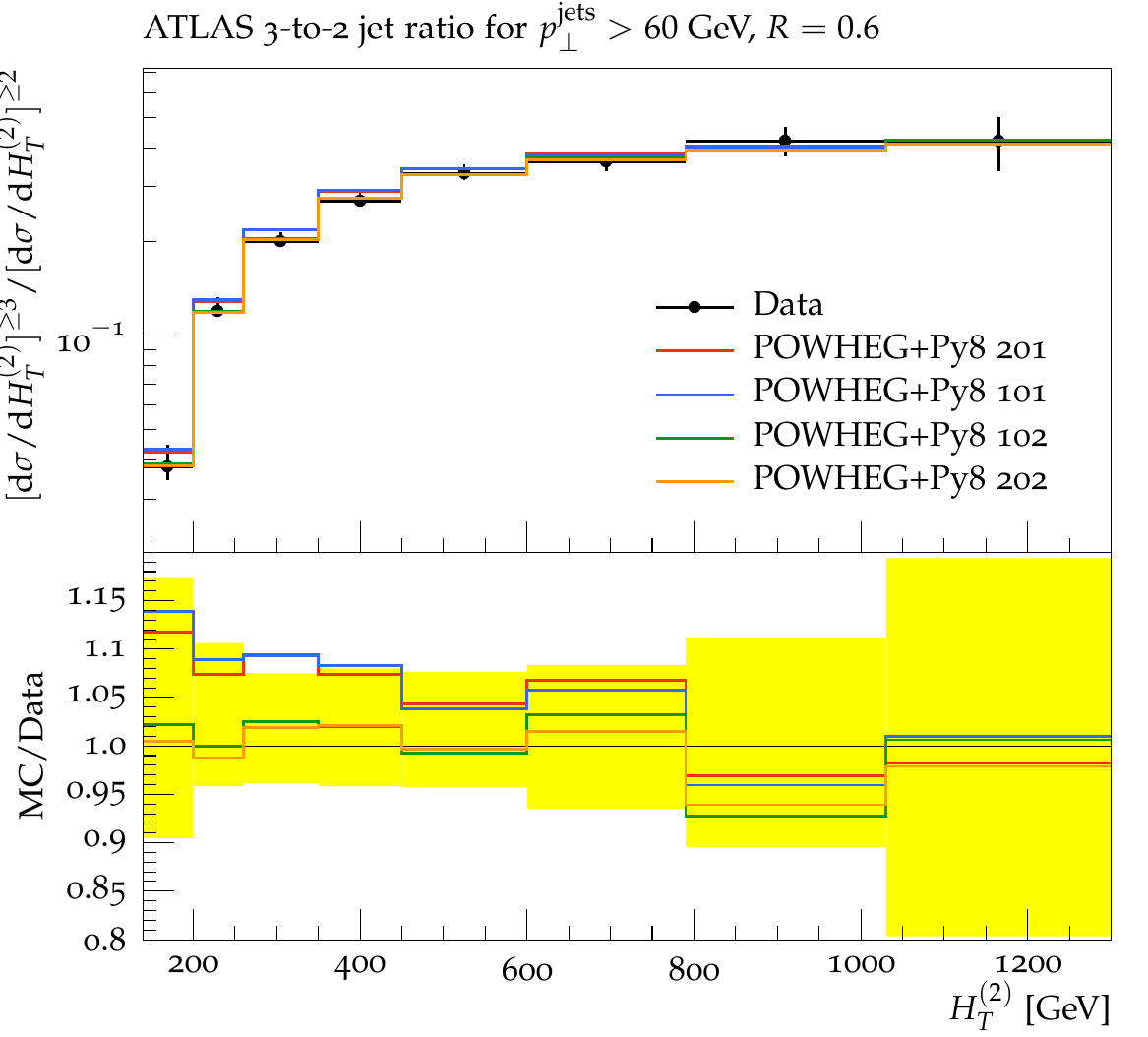}}\hfill
  \subcaptionbox{}{\img[0.48]{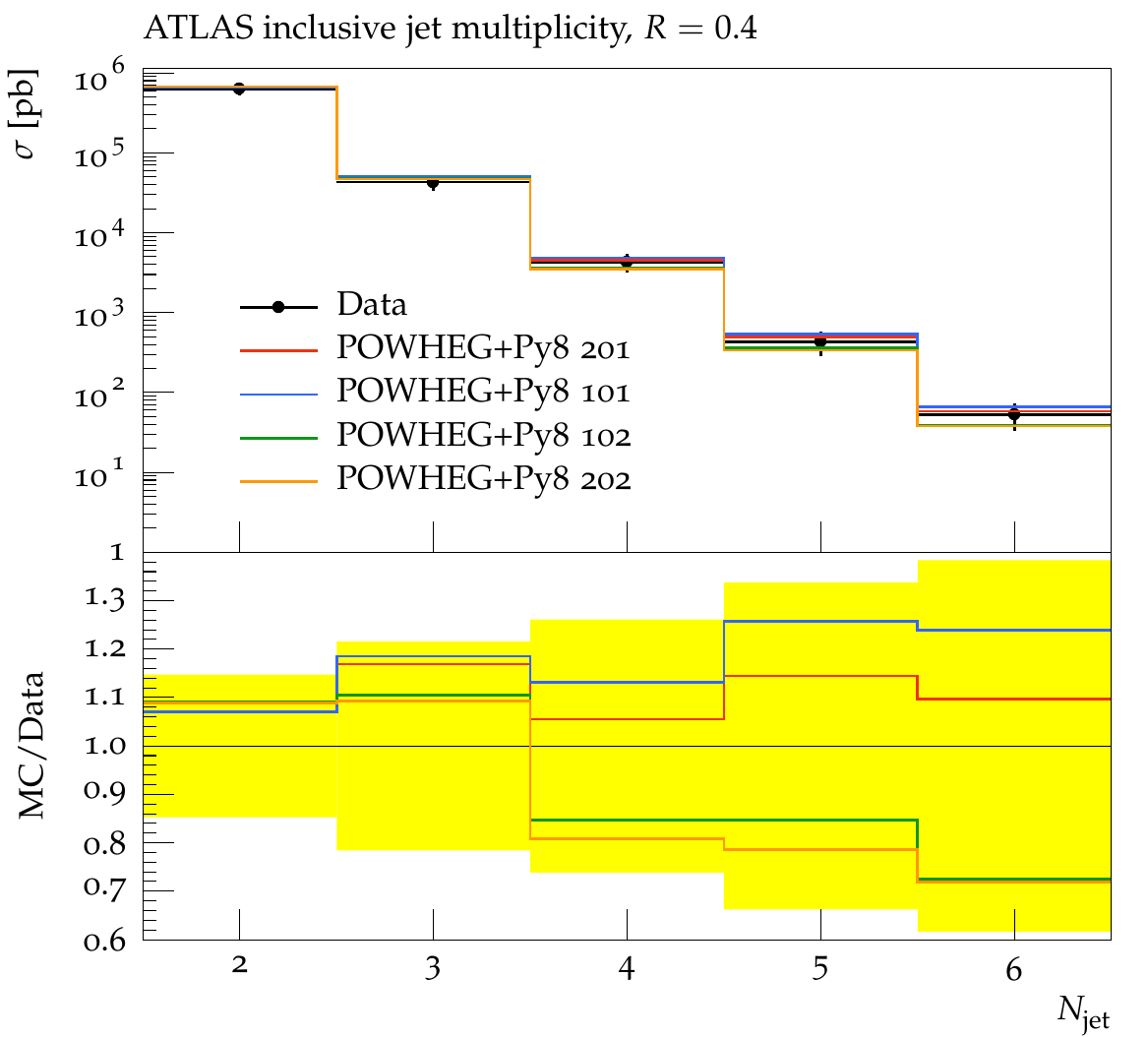}}
  \caption{Observables showing the effect of \pthard and \ptdef variations. The 3-digit tuple
    used in the plot legends represents the \PythiaEight $HED$ flag combination as
    described in the text.}
  \label{fig:ptharddefcmps2}
\end{figure}

\begin{figure}[tp]
  \centering
  \subcaptionbox{}{\img[0.48]{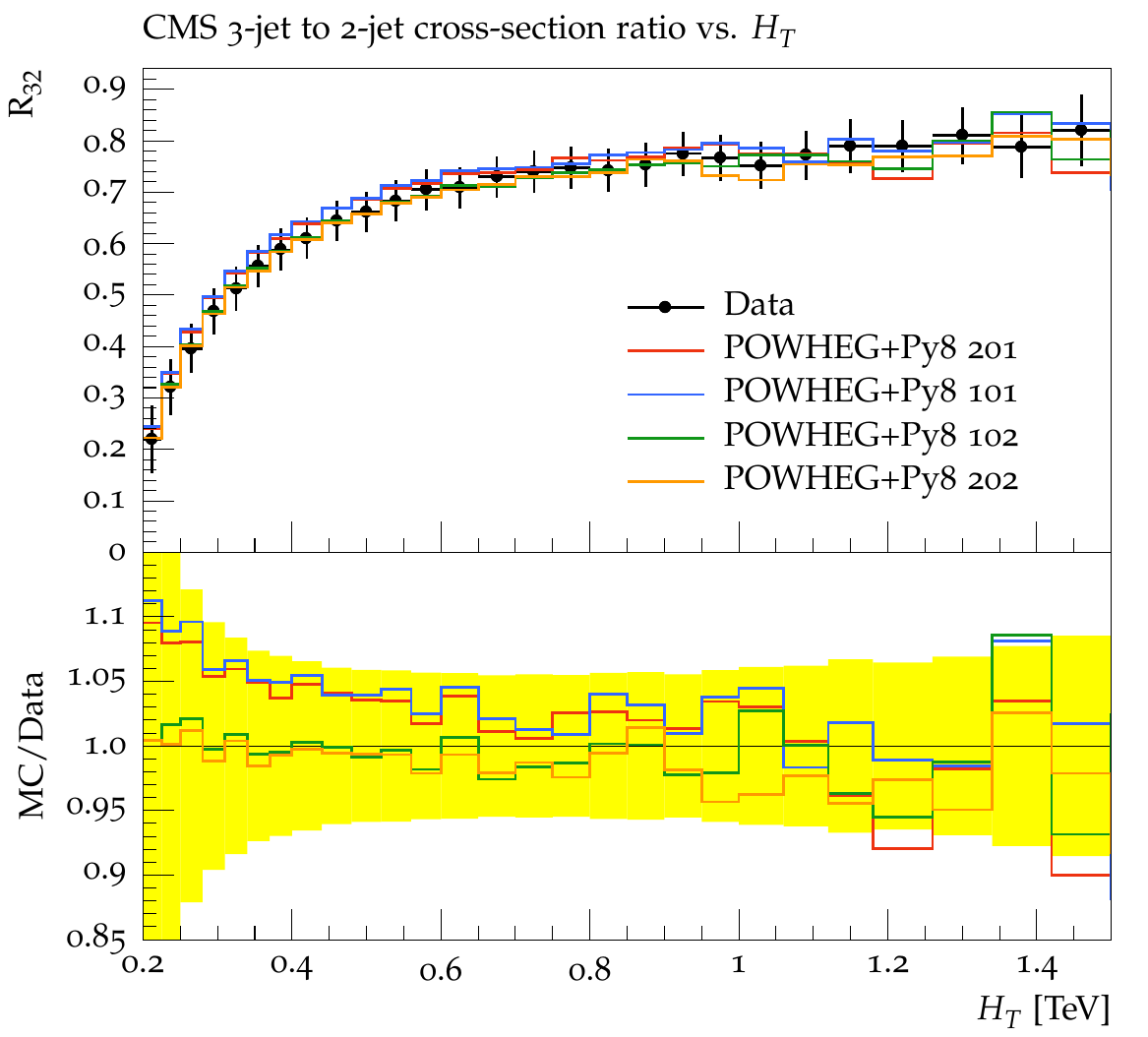}}\hfill
  \subcaptionbox{}{\img[0.48]{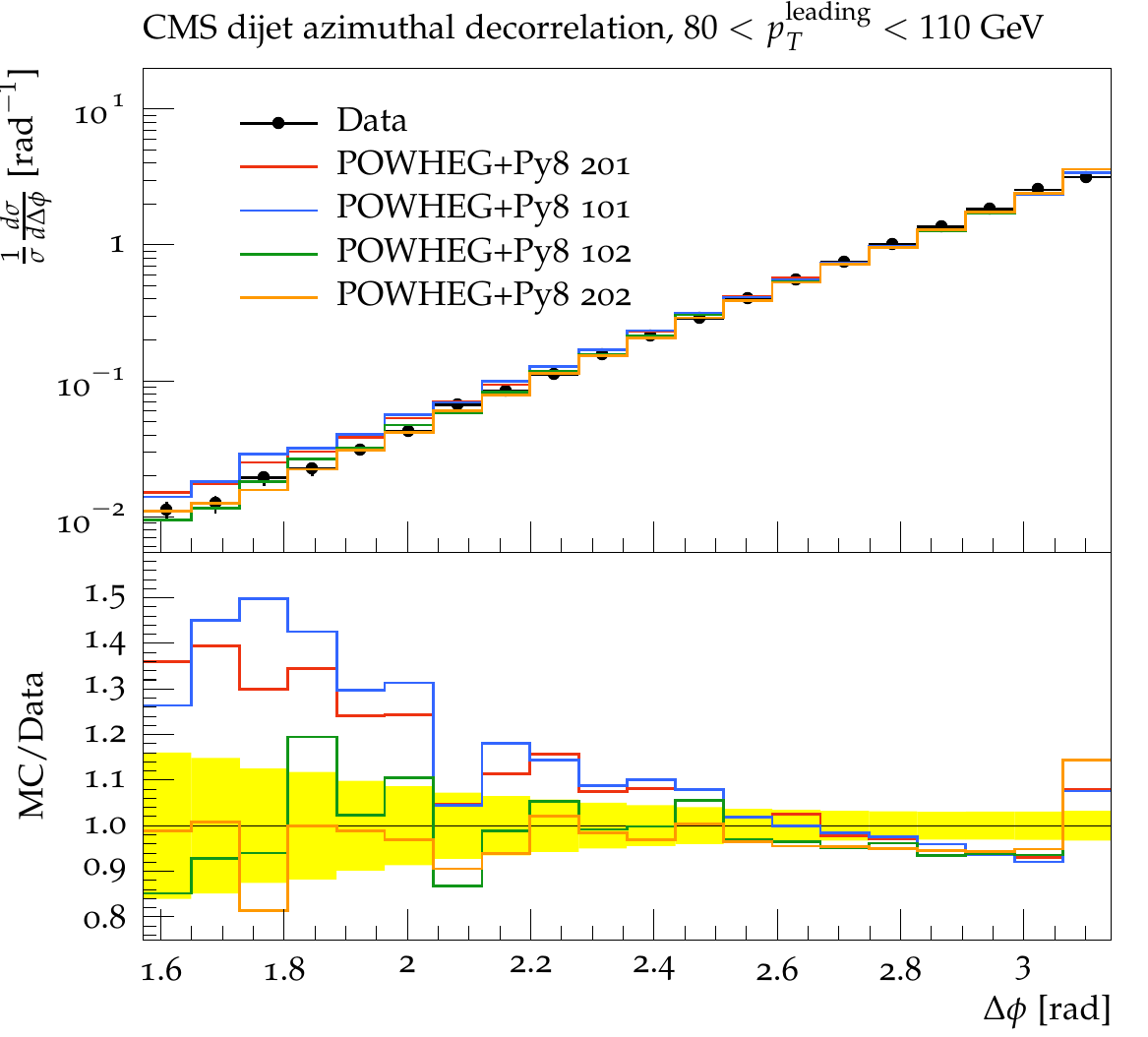}}\\[3ex]
  \subcaptionbox{}{\img[0.48]{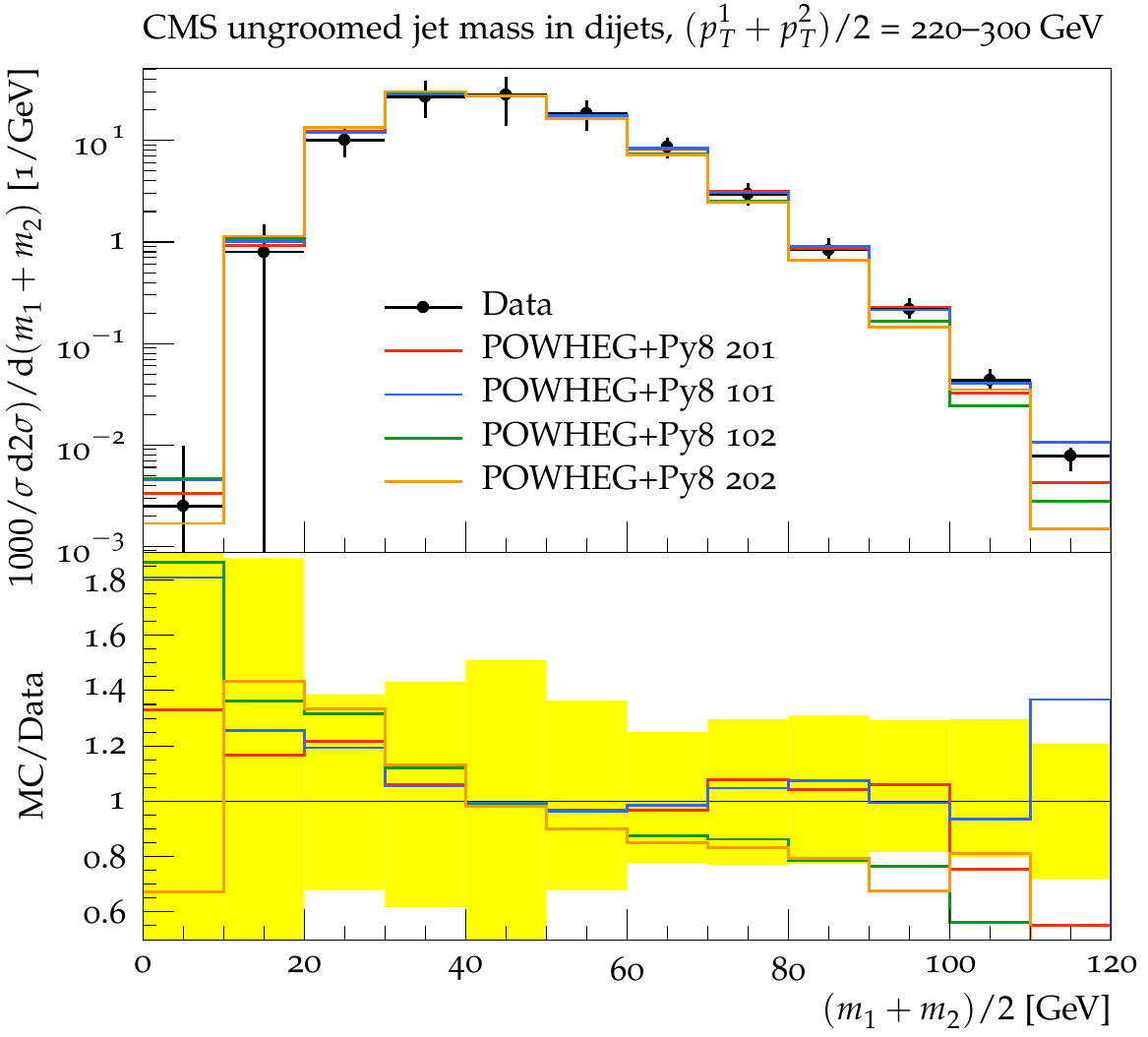}}\hfill
  \subcaptionbox{}{\img[0.48]{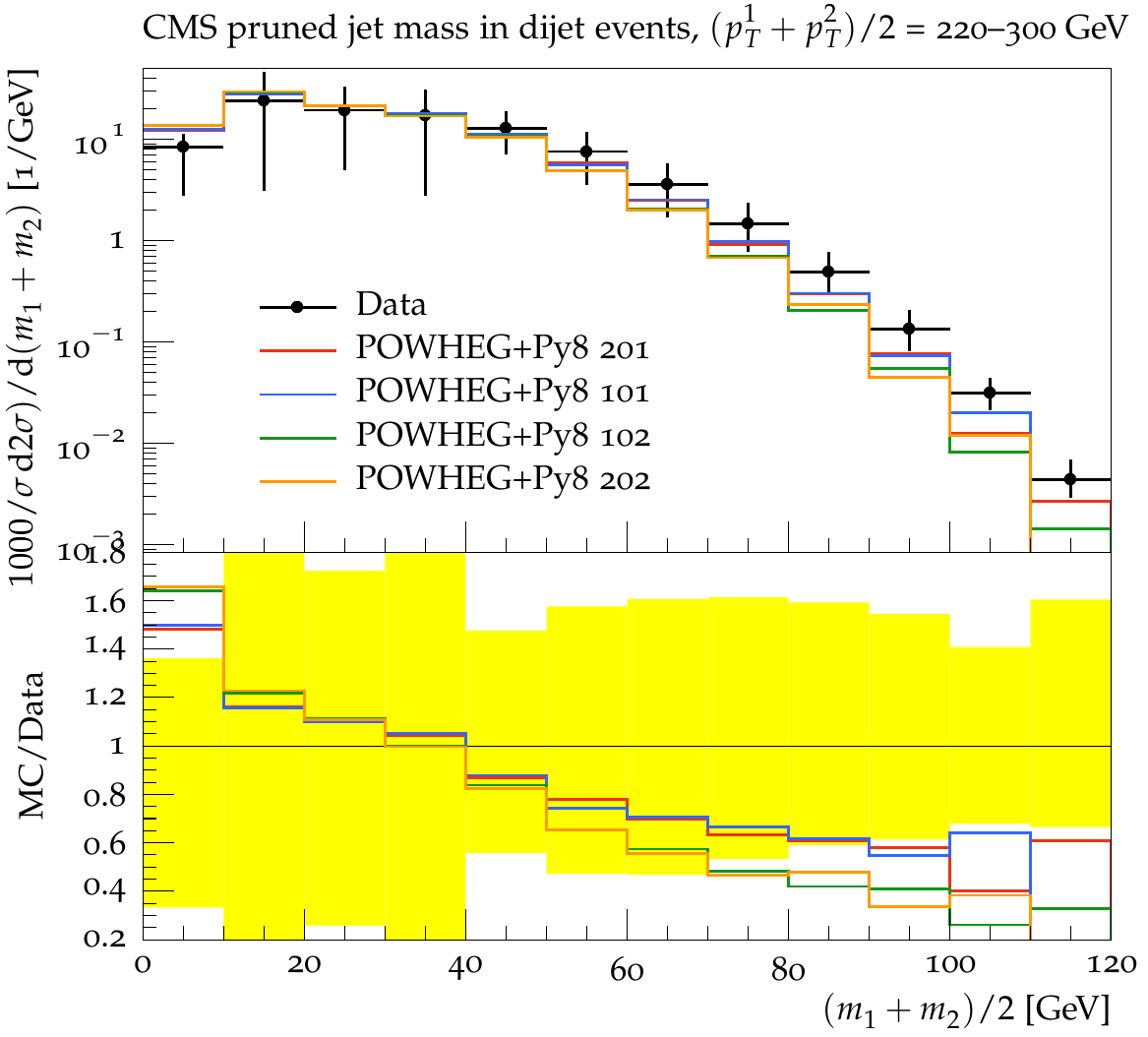}}
  \caption{Observables showing the effect of \pthard and \ptdef variations.  The
    3-digit tuple used in the plot legends represents the \PythiaEight $HED$
    flag combination as described in the text.}
  \label{fig:ptharddefcmps3}
\end{figure}

These variations are shown in Figures~\ref{fig:ptharddefcmps1}
to~\ref{fig:ptharddefcmps3}. The most distinctive feature of these histograms
shown in these figures is that there are two consistent groups of MC curves in
all cases: the 201 \& 101 combinations together, and the 102 \& 202 combinations
together. It is hence clear that the \ptdef mode (the 3rd component of the
$HED$ tuple) has a stronger effect than \pthard on these observables -- the
question is whether there is a clear preference for either grouping, and then
whether there is distinguishing power between the finer \pthard splitting within
the preferred group.

All the (ATLAS) plots in Figure~\ref{fig:ptharddefcmps1}
favour the $\ptdef=1$ grouping (the \PowhegBox \pt definition), in particular in
the dijet rapidity gap analysis where description of small rapidity gaps and
gaps between high mean-\pt dijet systems is significantly better than the
$\ptdef = 2$ option. The distinction is less significant -- within the
experimental error band -- for modelling of the 3-jet mass spectrum and for
larger rapidity gaps \& less energetic dijet systems.

In Figure~\ref{fig:ptharddefcmps2}, again the picture is less clear: neither
group of matching configurations describes transverse energy well as a function
of $|\eta|$, and both models converge to the same (poor) description at high
rapidity; the $\ptdef = 1$ configuration comes closer to the data at central
rapidities, falling just within the experimental error band, but $\ptdef = 2$
more closely matches the flatter \emph{shape} of the data. Interestingly, dijet
azimuthal decorrelation data prefers the $\ptdef = 2$ ``\Pythia \pt'' treatment
at the high-decorrelation (left-hand) end of the spectrum
but this region of the observable is expected to be affected by \emph{multiple}
extra emissions and hence the performance of the single-emission \Powheg scheme
is not clearly relevant. The 3-to-2 jet ratio also somewhat prefers the \Pythia
\pt scheme, but both schemes lie within the experimental error bars, as they do
for inclusive jet multiplicity modelling.

Finally, in Figure~\ref{fig:ptharddefcmps3} we again see mixed results: CMS'
characterisations of the 3-to-2 jet ratio and dijet decorrelations also prefer
$\ptdef = 2$, but the distribution of groomed jet masses exhibits a slight
preference for the $\ptdef = 1$ \PowhegBox \pt scheme.

The consistent trend throughout these observables is that the \PowhegBox \pt
definition produces more radiation, and hence larger $n$-jet masses, lower
rapidity gap fractions, more jet decorrelation, and more 3-jet events. Some of
these effects increase compatibility with data, while others prefer less
radiation, as produced by using the \Pythia \pt definition.

There is no clear ``best'' choice of \ptdef scheme from the available
observables, nor a consistent preference for either \pthard scheme within the
\ptdef groupings. \Pythia's default $HED = 201$ configuration is certainly
viable based on this comparison, but a \ptdef flip to the 202 configuration may be
a preferable choice if jet multiplicities are more important than their relative
kinematics (e.g. the multi-jet masses) for the application at hand.

We note that since the majority of these predictions fall within the
experimental uncertainties and are hence viable competing models, but there are
significant differences between them inside the experimental bands, \ptdef
variations may be a useful handle through which to estimate systematics
uncertainties in \Powheg+\Pythia matched simulations.




\section{\alphaS dependence of matched observables}

We conclude this short exploration of formal freedoms in \Powheg+\Pythia NLO
matching by studying the effect of different forms of the running strong
coupling $\alphaS(Q)$ in the parton shower. The \Powheg matrix element
necessarily uses an NLO \alphaS running with fixed value $\alphaS(M_Z) \sim 0.12$, for
consistency with the NLO PDF used in the calculation. It hence seems natural
that a parton shower matched to emissions using such an \alphaS should
itself\footnote{Or \emph{themselves}, since \Pythia's initial- and final-state
  showers can have separate couplings, and both participate in NLO matching.}
use such a coupling.

But a counter-argument is that the parton shower, as an iterated approximation
to the true QCD matrix elements for many-emission evolution, requires a large
\alphaS value to compensate for missing physics. This argument is formally
expressed in the proposal for ``CMW scaling''~\cite{Catani:1990rr} of a ``natural''
\alphaS, in which its divergence scale $\Lambda_\text{QCD}$ is scaled up by an
$N_F$-dependent factor between 1.5 and 1.7. Such a scaling increases the
effective $\alphaS(M_Z)$ of a ``bare'' NLO strong coupling.  And empirically a
fairly large \alphaS \emph{is} found to be preferred in MC
tuning~\cite{Buckley:2009bj,ATL-PHYS-PUB-2014-021,Skands:2014pea}, in particular for description of final-state
effects like jet shapes and masses. These large couplings can become as dramatic
as $\alphaS(M_Z) \sim 0.14$ without any ill effect on such observables. Whether
one- or two-loop \alphaS running is more appropriate in \Powheg matching is also
an open question with arguments possible in both directions.

\begin{figure}[tp]
  \centering
  \subcaptionbox{}{\img[0.48]{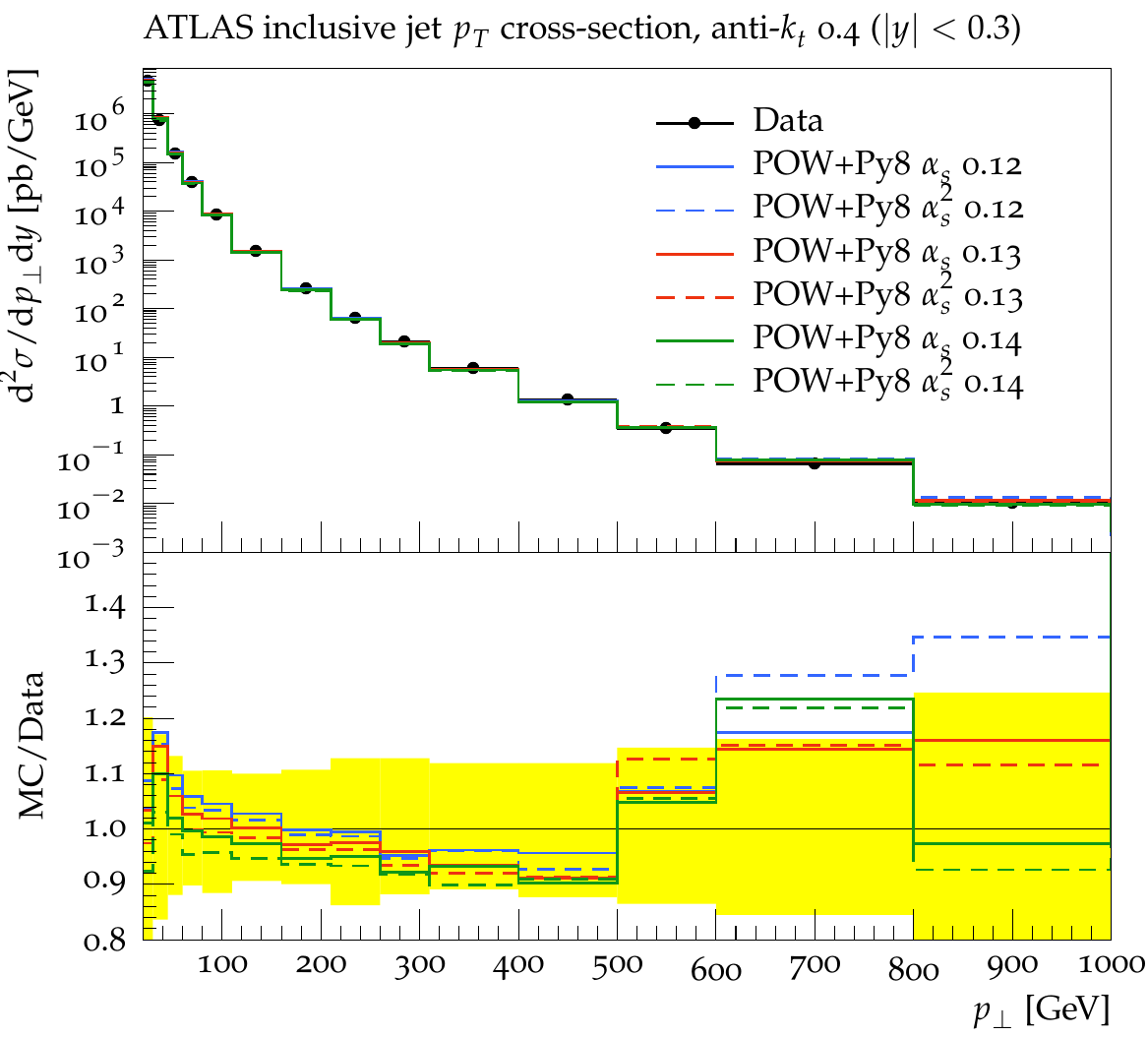}}\hfill
  \subcaptionbox{}{\img[0.48]{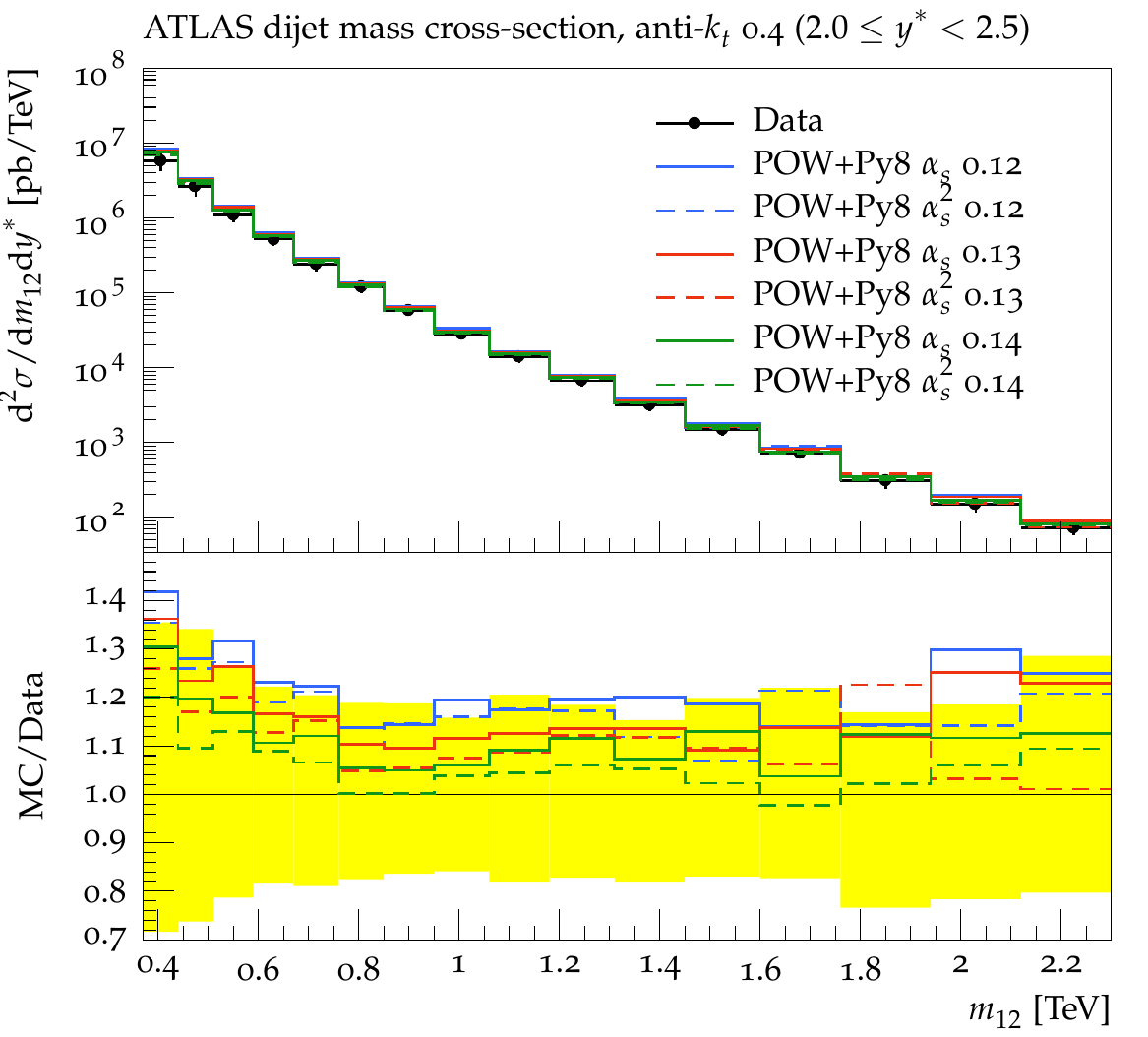}}\\[3ex]
  \subcaptionbox{}{\img[0.48]{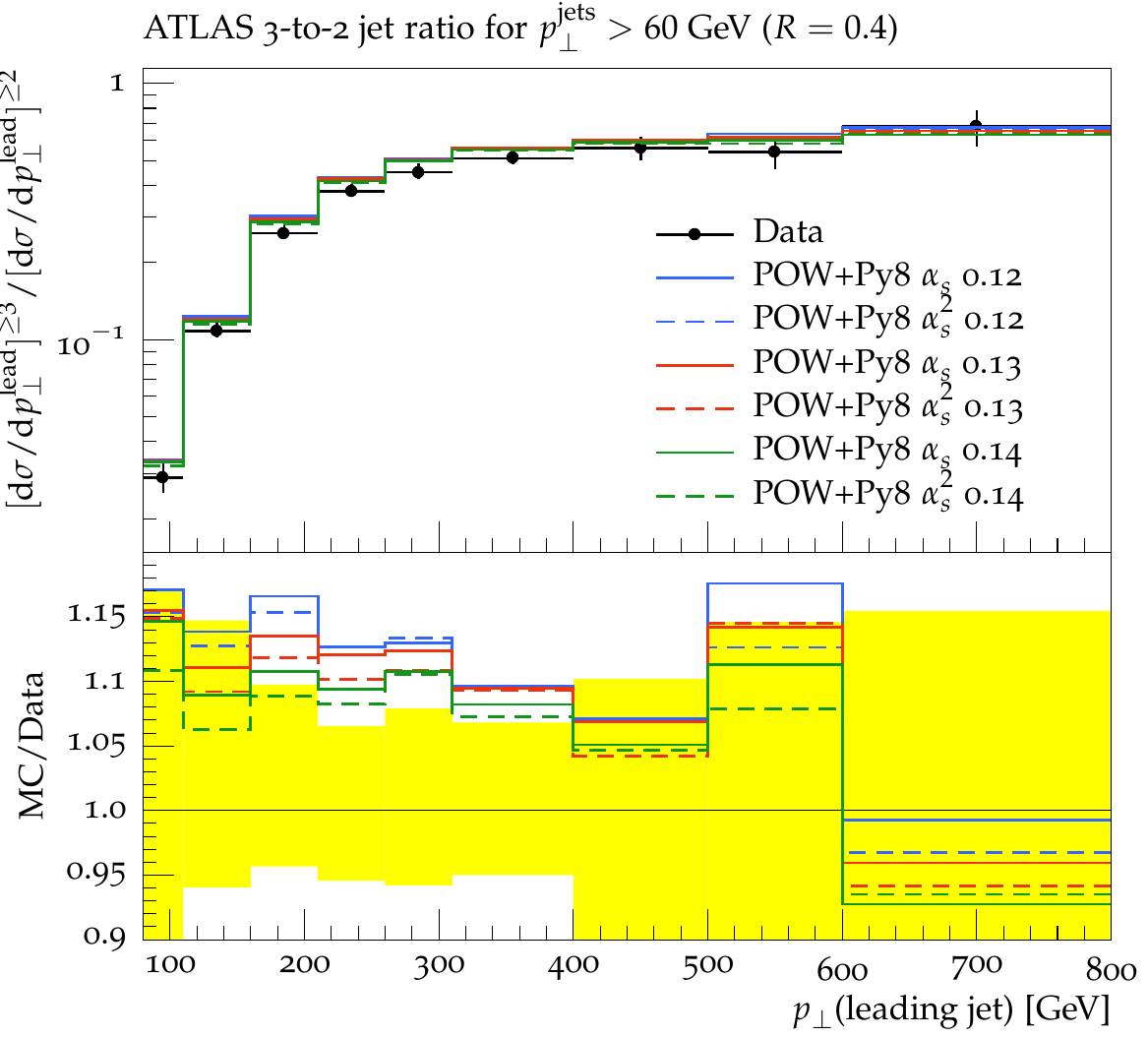}}\hfill
  \subcaptionbox{}{\img[0.48]{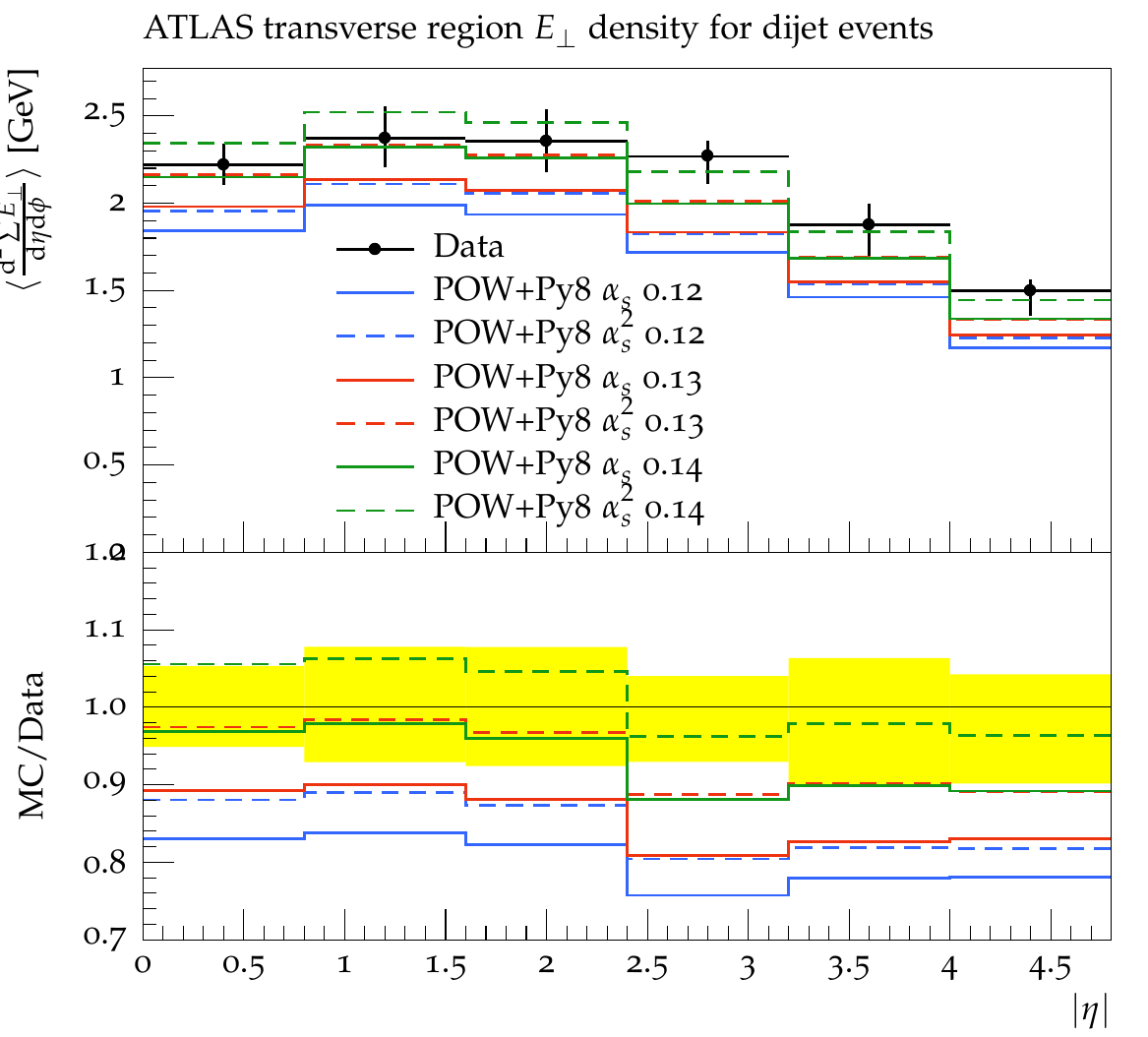}}
  \caption{Observables showing the effect of \alphaS variation, in tandem for
    \Pythia's ISR and FSR parton showers. The legend indicates the values of
    $\alphaS(M_Z)$ used to fix the running coupling, and whether a 1-loop or
    2-loop $\beta$-function is being used.}
  \label{fig:alphascmps1}
\end{figure}

\begin{figure}[p]
  \thisfloatpagestyle{empty}
  \centering
  \vspace*{-1.8em}
  \subcaptionbox{}{\img[0.44]{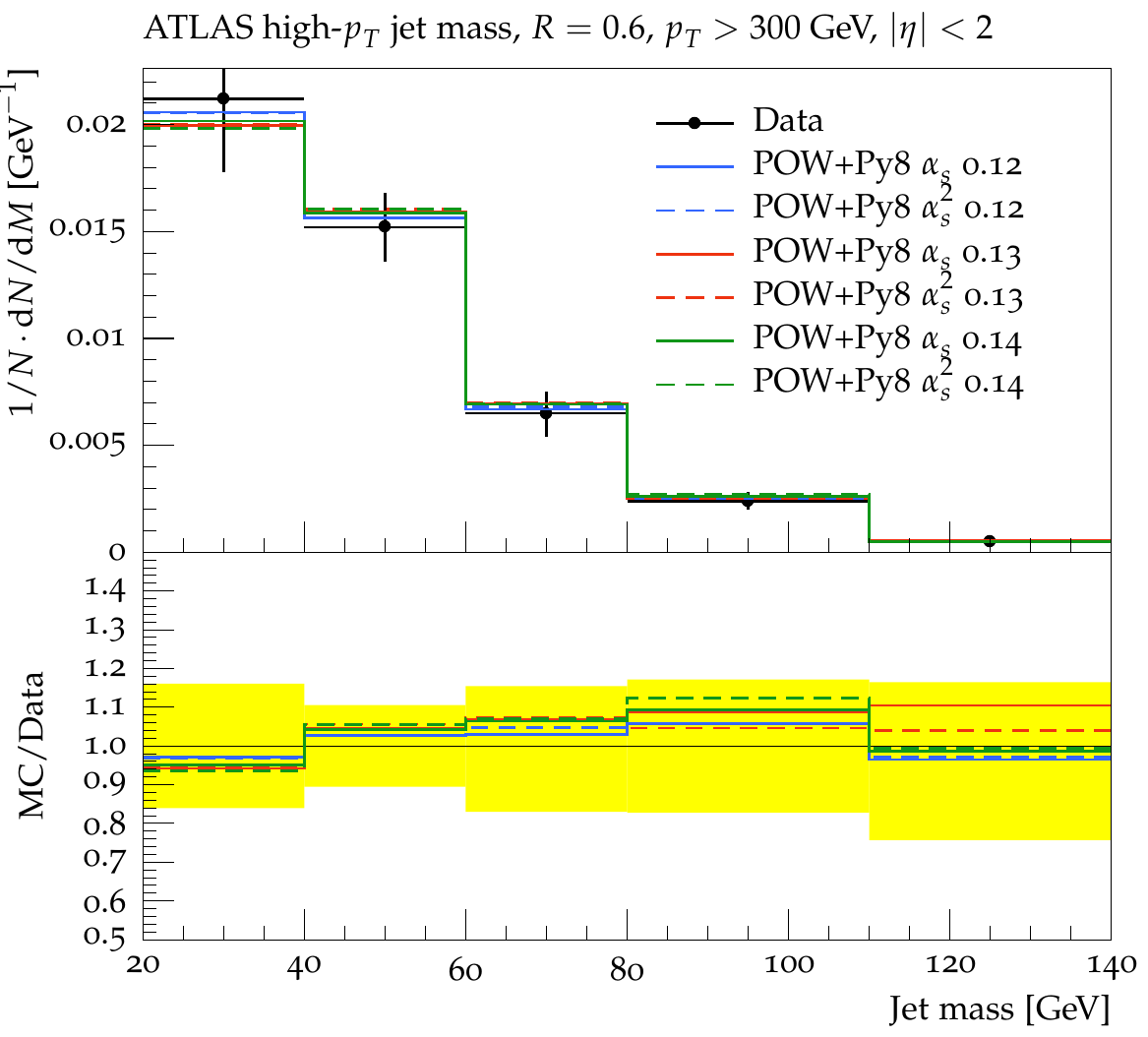}}\hfill
  \subcaptionbox{}{\img[0.44]{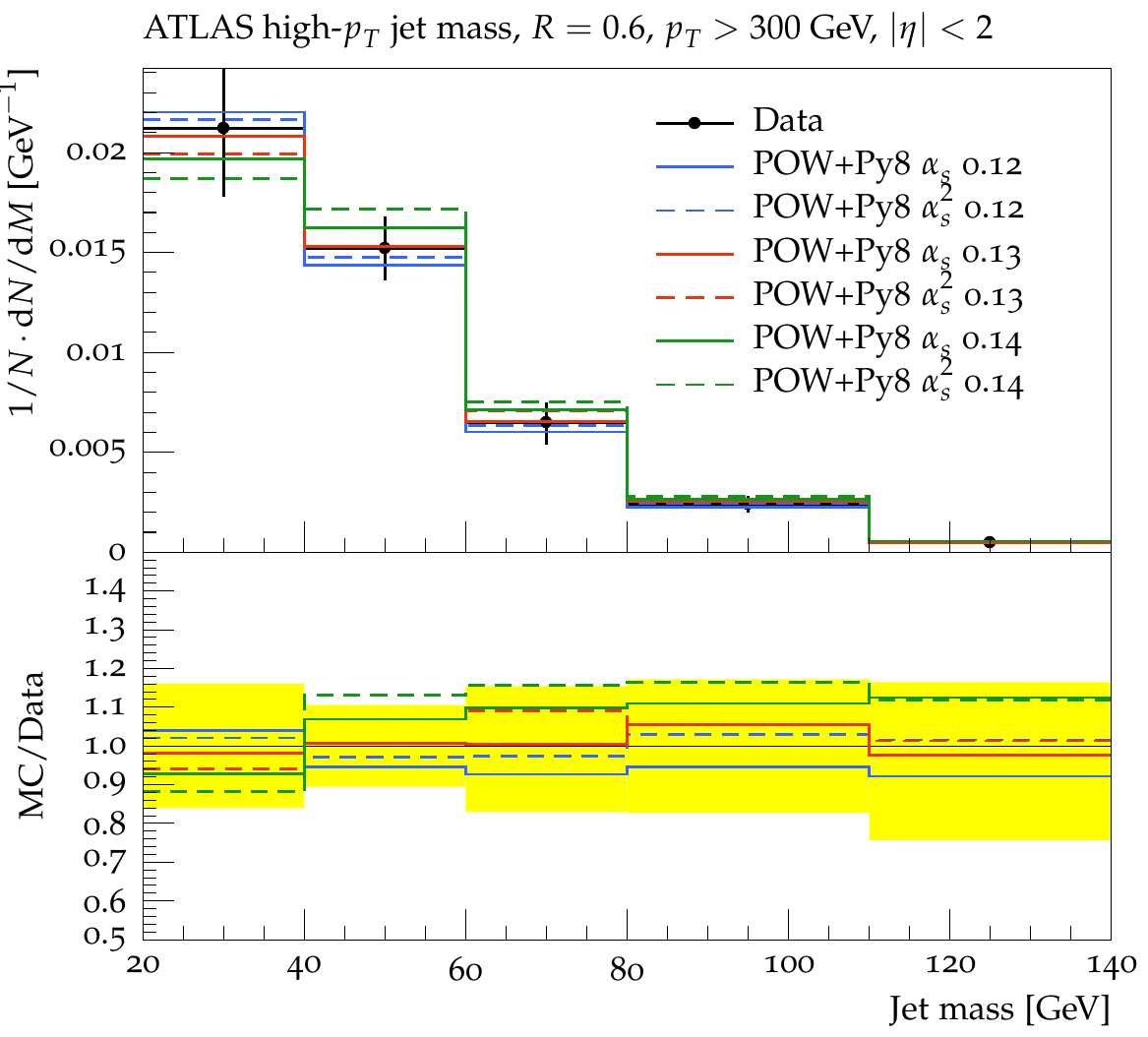}}\\[2ex]
  \subcaptionbox{}{\img[0.44]{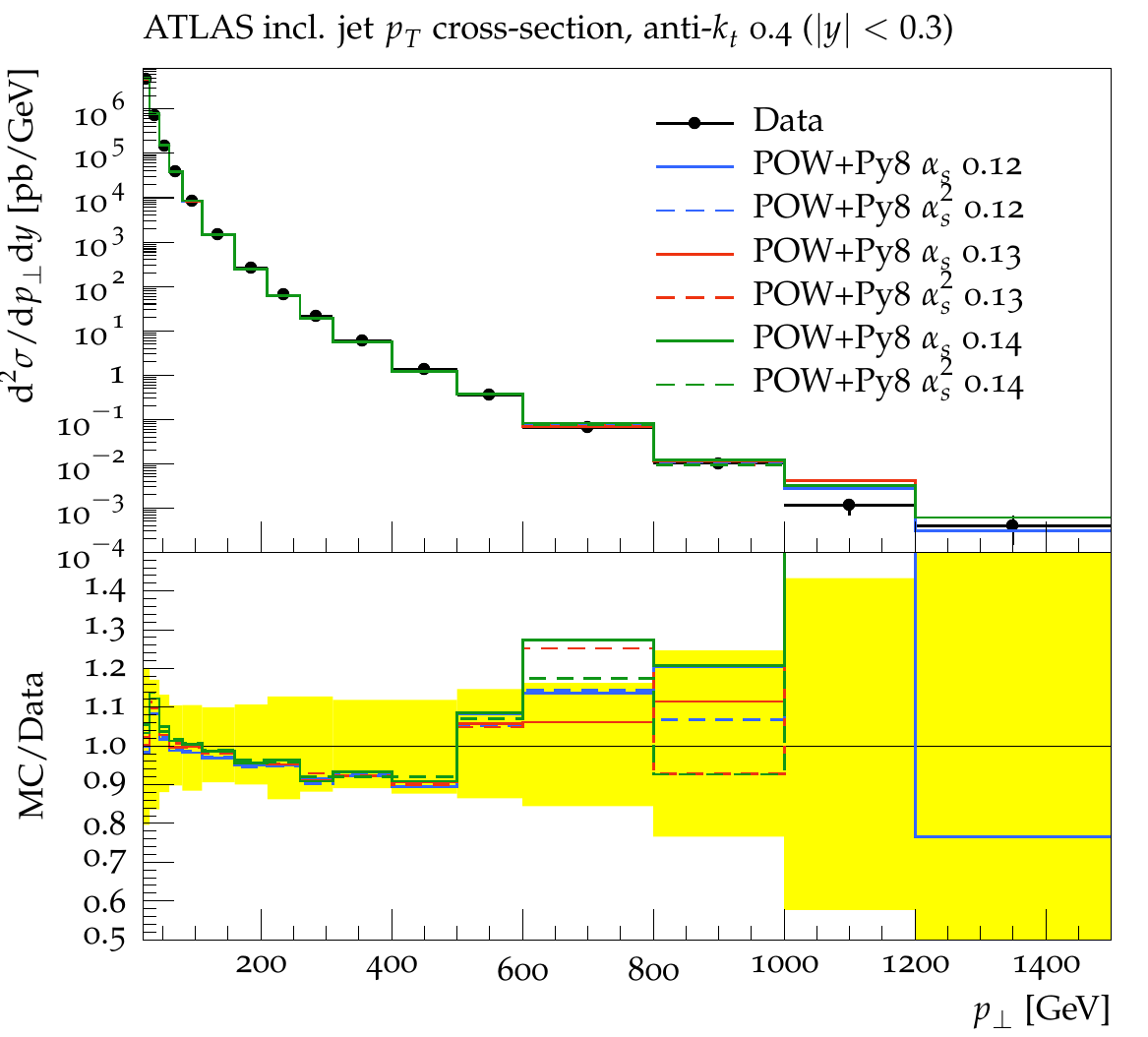}}\hfill
  \subcaptionbox{}{\img[0.44]{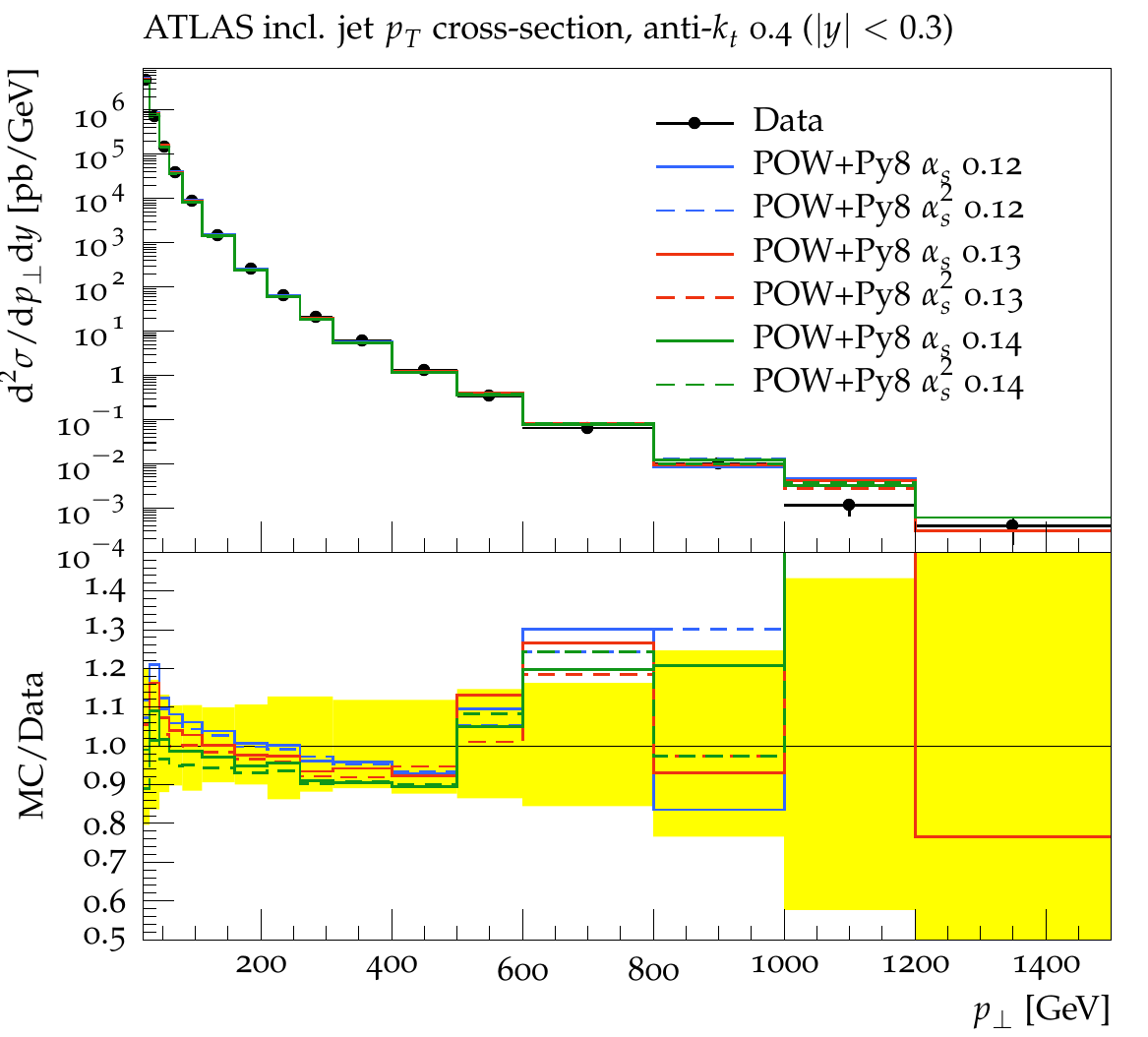}}\\[2ex]
  \subcaptionbox{}{\img[0.44]{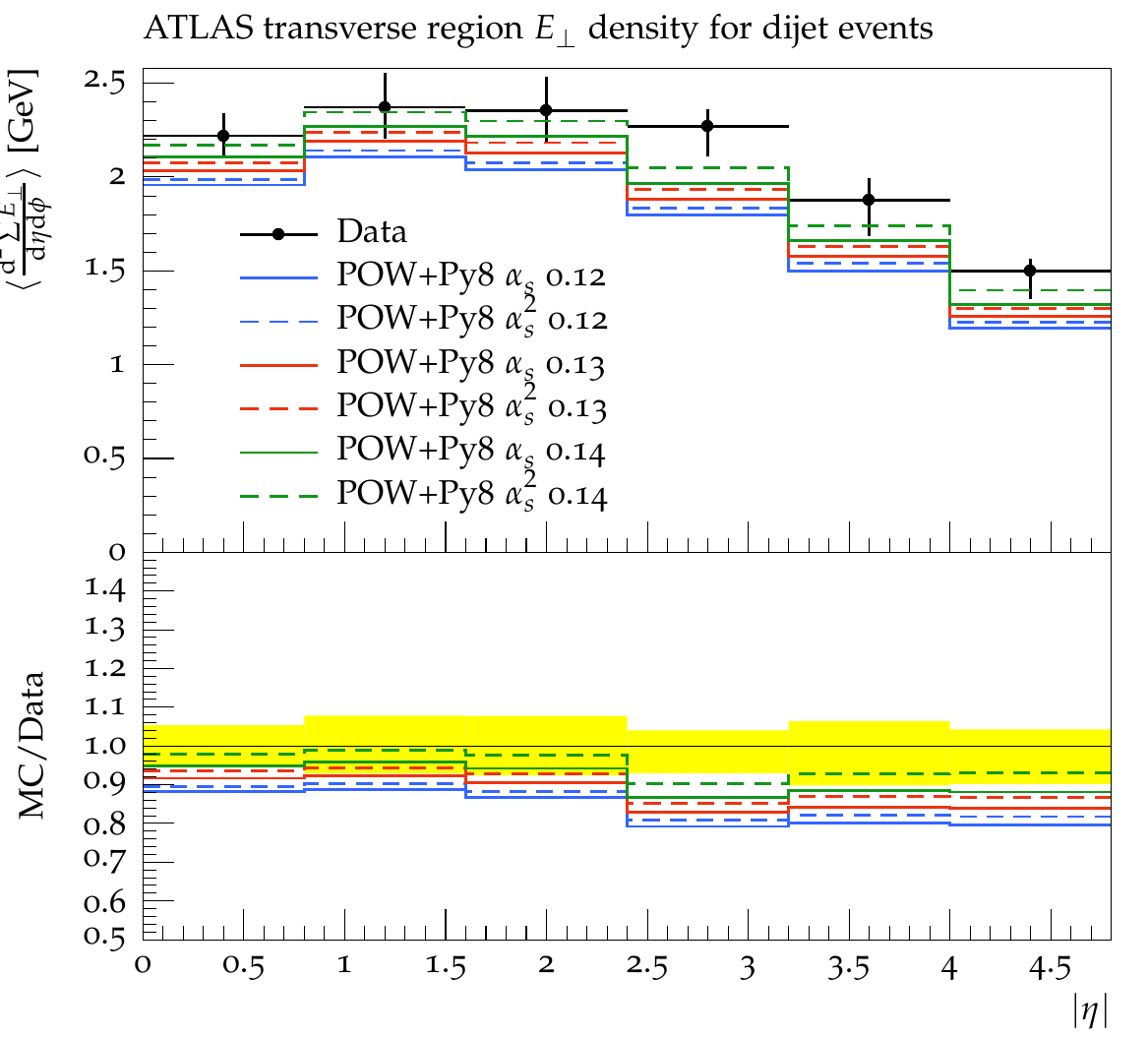}}\hfill
  \subcaptionbox{}{\img[0.44]{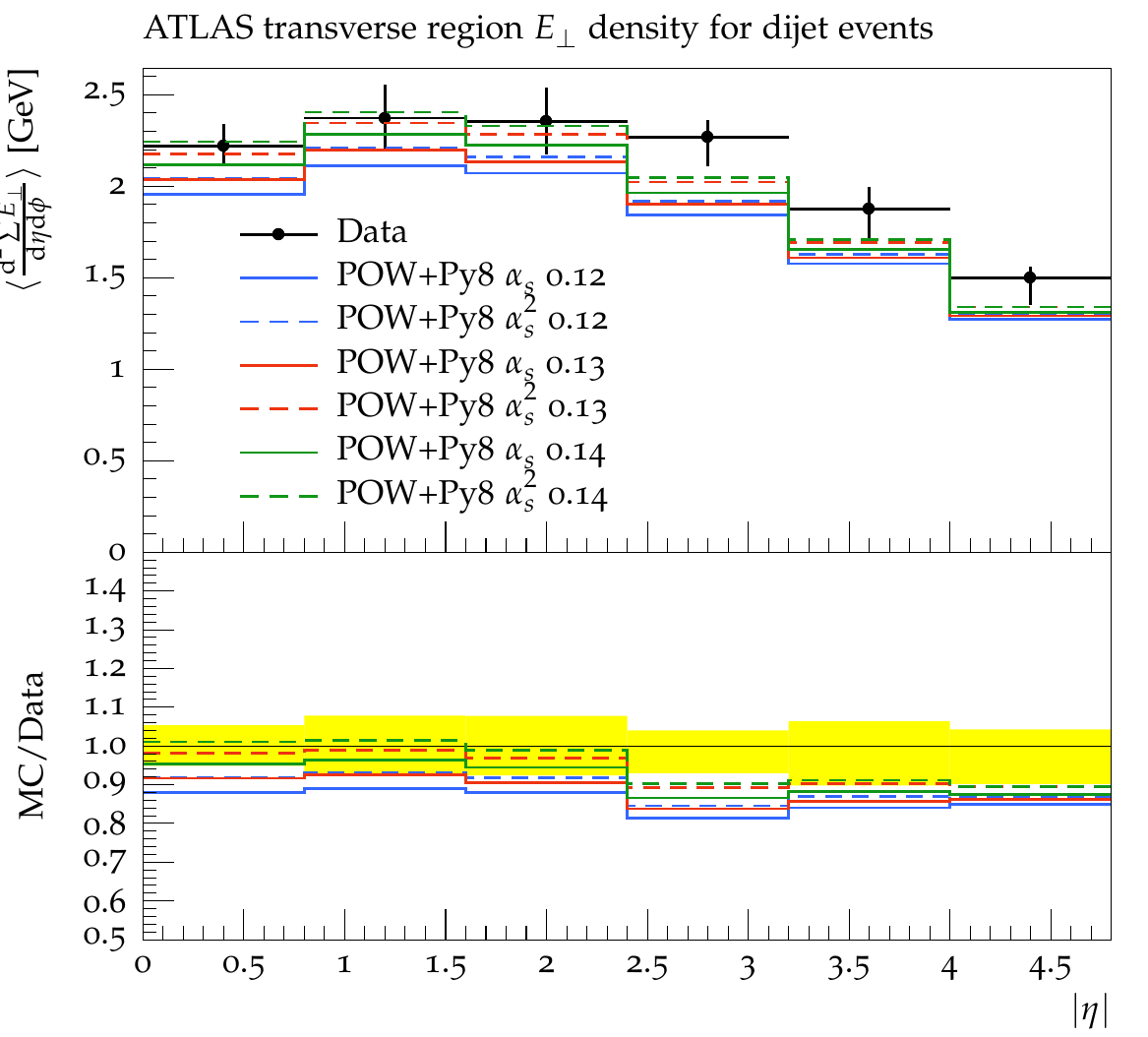}}
  \caption{Observables showing the effect of \alphaS variation.  The legend
    indicates the values of $\alphaS(M_Z)$ used to fix the running coupling, and
    whether a 1-loop or 2-loop $\beta$-function is being used. The left-hand
    column is for variation of the ISR shower \alphaS only, and the right-hand
    column for variation of the FSR \alphaS only. Due to \Powheg matching, the
    ISR shower has essentially no effect, even on ``ISR observables''.}
  \label{fig:alphascmps2}
\end{figure}

Since several arguments can be made for using $\alphaS(M_Z)$ values between 0.12
and 0.14 (respectively, consistency, CMW, and naked pragmatism), rather than
explicitly perform CMW or similar scalings, we have explored 6 configurations
with $\alphaS(M_Z) \in \{ 0.12, 0.13, 0.14 \}$ ("NLO-like", "LO-like" and
"enhanced LO", respectively), and 1-loop and 2-loop running in each. Observables
demonstrating these variations, each using 10M events with identical \alphaS
configurations in ISR and FSR showers, can be found in Figures~\ref{fig:alphascmps1}
and~\ref{fig:alphascmps2}.



In Figure~\ref{fig:alphascmps1} all variations are grouped tightly within the
experimental uncertainty band, but a clear trend of higher \alphaS producing
lower cross-sections is visible. A similar effect is visible in the dijet
cross-section as a function of dijet mass, with all bins favouring a strong
$\alphaS(M_Z) = 0.14$ with 2-loop running. Similar effects were seen for the
equivalent CMS jet \pt spectra. The same strong \alphaS settings are slightly
but consistently preferred by the 3/2 jet ratio data, and the transverse energy
flow also favours a large coupling, although not quite so extreme: either 1-loop
$\alphaS(M_Z) = 0.14$ or 2-loop $\alphaS(M_Z) = 0.13$ work best.  In all these
observables, the steps of 0.1 in $\alphaS(M_Z)$ have a similar magnitude of
effect to the switches between 1-loop (``LO'') and 2-loop (``NLO'') running.



This is interesting: there is a fairly strong preference across much
``ISR-influenced'' data for a large shower coupling with
$\alphaS(M_Z) \sim 0.14$, even in \Powheg-matched simulation. More generally,
\alphaS variations on this scale again provide useful coverage of experimental
error bands and are hence a useful systematic handle on \Powheg simulation.

But FSR-dominated jet shapes and masses usually have a strong preference (if
only due to the narrowness of jet shape experimental uncertainty bands) for a
slightly smaller $\alphaS(M_Z) \sim 0.13$. While matching might imply a need for
consistency between ISR and FSR shower configurations, we are already operating
in a regime beyond the formal accuracy of the method, and it is intriguing to
see whether a hybrid setup with $\alphaS(M_Z) \sim 0.14$ in the initial-state
shower and $\alphaS(M_Z) \sim 0.13$ in the final-state shower would give the
best of both worlds across ``ISR'' and ``FSR'' observables. We do this by
separately setting one shower $\alphaS(M_Z)$ to the default 0.13 and varying the
other, and vice~versa, with results shown in Figure~\ref{fig:alphascmps2}.

The results are illuminating: while as usual changes in the ISR shower have
little effect on jet masses, they also have no effect on jet \pt -- a classic
``ISR'' observable. This pattern was repeated for all ``ISR observables'', with
virtually zero effect of the ISR shower coupling in all cases. The only variable
to exhibit ISR shower sensitivity is the soft-QCD-dominated transverse energy
flow, which is equally determined by ISR and FSR showering. While na\"ively
counter-intuitive, this makes perfect sense: the \Powheg hard process is
providing almost all the ISR effects, and leaving little phase space for the
initial-state shower to produce effects on hard jet observables. These plots
hence illustrate that the FSR shower configuration can have significant effects
upon \Powheg matching, and that there is a tension between the higher FSR
\alphaS desired to describe 3/2 jet ratios and multi-jet mass spectra, and the
slightly lower values to describe intra-jet effects such as single-jet masses.









\section{Summary}

In this note we have explored several freedoms in \PythiaEight's machinery for
matching parton showers to hard process partonic events generated by the
\PowhegBox MC generator. These have included discrete options for defining and
calculating the \Powheg emission vetoing scale, and the continuous freedom to
vary the strong coupling in \Pythia's parton showers.

All these freedoms are permitted within the fixed-order accuracy of the \Powheg
method, but it is clear that some will be better choices than others. We have
hence considered the full set of 7~\TeV $pp$ data analyses available from the
ATLAS and CMS collaborations, via the Rivet analysis system, to determine both
whether there is an unambiguously preferred interpretation of the \Powheg
matching scheme, and whether ``reasonable'' variations on the nominal scheme can
be used to estimate systematic uncertainties from the matching, to be combined
with fixed-order scale uncertainties.

We conclude that the default \PythiaEight ``\kbd{main31}'' matching
configuration is a viable matching scheme but it is not unique in this. The
calculation of the ``hardness'' of a proposed parton shower emission must
necessarily be chosen to match\footnote{As closely as possible -- this is itself
  not an unambiguous choice, maybe suggesting a further route for systematics
  exploration.} that used in the \Powheg hard process, on pain of huge data/MC
discrepancies -- but there is much less clarity about the choice of \pt
definition to be used in the scale calculation and (less importantly) the
approach taken to recalculate the \Powheg ME event's veto scale. The data
suggests that in observables concerned more with jet multiplicity than
kinematics, an alternative \pt definition may perform better, and that in
general variations of \pt definition may be a useful handle on \Powheg--\Pythia
matching uncertainty.

Variations of the strong coupling in the \PythiaEight parton showers between
``NLO-like'' and ``LO- or LL-like'' $\alphaS(M_Z)$ values also gave good
coverage of the experimental data uncertainties, and provide an alternative
route for systematics evaluation. Interestingly, the \Powheg matching has been
seen to almost completely eliminate sensitivity to the \Pythia initial-state
parton shower in inter-jet observables like jet \pt and multi-jet masses --
the final-state shower, often caricatured as only affecting intra-jet
observables like jet shapes and masses, is responsible for almost all shower
effects on observables, even those dominated by ISR. Obviously this simply
reflects the fact that \Powheg vetoing constrains the emission phase space of
the initial-state shower far more than the final-state one, but it may have
implications for NLO-matched shower generator tuning, e.g. using the FSR
coupling to optimise ``ISR observables'' and the ISR shower freedom to purely
improve the description of soft effects like underlying event and transverse
energy flow.

\subsection*{Acknowledgements}

This work was supported by the European Union Marie Curie Research Training
Network MCnetITN, under contract PITN-GA-2012-315877.  Our thanks to
Stefan~Prestel for several useful discussions, insights into the \Pythia
matching options, and admirable tenacity in awaiting the overdue completion of
this paper!

\bibliographystyle{JHEP}
\bibliography{refs}

\providecommand{\href}[2]{#2}\begingroup\raggedright\begin{thebibliography}{10}

\bibitem{Mangano:2002ea}
M.~L. Mangano, M.~Moretti, F.~Piccinini, R.~Pittau, and A.~D. Polosa, {\em
  {\textsc{Alpgen}, a generator for hard multiparton processes in hadronic
  collisions}\/},  \href{http://dx.doi.org/10.1088/1126-6708/2003/07/001}{JHEP
  {\bf 07} (2003)  001},
\href{http://arxiv.org/abs/hep-ph/0206293}{{\tt arXiv:hep-ph/0206293
  [hep-ph]}}.

\bibitem{Alwall:2011uj}
J.~Alwall, M.~Herquet, F.~Maltoni, O.~Mattelaer, and T.~Stelzer, {\em {MadGraph
  5 : Going Beyond}\/},  \href{http://dx.doi.org/10.1007/JHEP06(2011)128}{JHEP
  {\bf 06} (2011)  128},
\href{http://arxiv.org/abs/1106.0522}{{\tt arXiv:1106.0522 [hep-ph]}}.

\bibitem{Gleisberg:2008ta}
T.~Gleisberg, S.~Hoeche, F.~Krauss, M.~Schonherr, S.~Schumann, F.~Siegert, and
  J.~Winter, {\em {Event generation with SHERPA 1.1}\/},
  \href{http://dx.doi.org/10.1088/1126-6708/2009/02/007}{JHEP {\bf 02} (2009)
  007},
\href{http://arxiv.org/abs/0811.4622}{{\tt arXiv:0811.4622 [hep-ph]}}.

\bibitem{Frixione:2007vw}
S.~Frixione, P.~Nason, and C.~Oleari, {\em {Matching NLO QCD computations with
  parton shower simulations: the \Powheg method}\/},
  \href{http://dx.doi.org/10.1088/1126-6708/2007/11/070}{JHEP {\bf 11} (2007)
  070},
\href{http://arxiv.org/abs/0709.2092}{{\tt arXiv:0709.2092 [hep-ph]}}.

\bibitem{Frixione:2002bd}
S.~Frixione and B.~R. Webber, {\em {The MC@NLO event generator}\/},
\href{http://arxiv.org/abs/hep-ph/0207182}{{\tt arXiv:hep-ph/0207182
  [hep-ph]}}.

\bibitem{Alwall:2014hca}
J.~Alwall, R.~Frederix, S.~Frixione, V.~Hirschi, F.~Maltoni, O.~Mattelaer,
  H.~S. Shao, T.~Stelzer, P.~Torrielli, and M.~Zaro, {\em {The automated
  computation of tree-level and next-to-leading order differential cross
  sections, and their matching to parton shower simulations}\/},
  \href{http://dx.doi.org/10.1007/JHEP07(2014)079}{JHEP {\bf 07} (2014)  079},
\href{http://arxiv.org/abs/1405.0301}{{\tt arXiv:1405.0301 [hep-ph]}}.

\bibitem{Frederix:2012ps}
R.~Frederix and S.~Frixione, {\em {Merging meets matching in MC@NLO}\/},
  \href{http://dx.doi.org/10.1007/JHEP12(2012)061}{JHEP {\bf 12} (2012)  061},
\href{http://arxiv.org/abs/1209.6215}{{\tt arXiv:1209.6215 [hep-ph]}}.

\bibitem{Sjostrand:2014zea}
T.~Sjostrand, S.~Ask, J.~R. Christiansen, R.~Corke, N.~Desai, P.~Ilten,
  S.~Mrenna, S.~Prestel, C.~O. Rasmussen, and P.~Z. Skands, {\em {An
  Introduction to PYTHIA 8.2}\/},
  \href{http://dx.doi.org/10.1016/j.cpc.2015.01.024}{Comput. Phys. Commun. {\bf
  191} (2015)  159--177},
\href{http://arxiv.org/abs/1410.3012}{{\tt arXiv:1410.3012 [hep-ph]}}.

\bibitem{Alioli:2010xa}
S.~Alioli, K.~Hamilton, P.~Nason, C.~Oleari, and E.~Re, {\em {Jet pair
  production in \Powheg}\/},
  \href{http://dx.doi.org/10.1007/JHEP04(2011)081}{JHEP {\bf 04} (2011)  081},
\href{http://arxiv.org/abs/1012.3380}{{\tt arXiv:1012.3380 [hep-ph]}}.

\bibitem{Buckley:2010ar}
A.~Buckley, J.~Butterworth, L.~Lonnblad, D.~Grellscheid, H.~Hoeth, J.~Monk,
  H.~Schulz, and F.~Siegert, {\em {Rivet user manual}\/},
  \href{http://dx.doi.org/10.1016/j.cpc.2013.05.021}{Comput. Phys. Commun. {\bf
  184} (2013)  2803--2819},
\href{http://arxiv.org/abs/1003.0694}{{\tt arXiv:1003.0694 [hep-ph]}}.

\bibitem{Alwall:2006yp}
J.~Alwall et al., {\em {A standard format for Les Houches event files}\/},
  \href{http://dx.doi.org/10.1016/j.cpc.2006.11.010}{Comput. Phys. Commun. {\bf
  176} (2007)  300--304},
\href{http://arxiv.org/abs/hep-ph/0609017}{{\tt arXiv:hep-ph/0609017
  [hep-ph]}}.

\bibitem{Dobbs:2001ck}
M.~Dobbs and J.~B. Hansen, {\em {The HepMC C++ Monte Carlo event record for
  High Energy Physics}\/},
\href{http://dx.doi.org/10.1016/S0010-4655(00)00189-2}{Comput. Phys. Commun.
  {\bf 134} (2001)  41--46}.

\bibitem{Aad:2014rma}
{ATLAS} Collaboration, G.~Aad et al., {\em {Measurement of three-jet production
  cross-sections in $pp$ collisions at 7 TeV centre-of-mass energy using the
  ATLAS detector}\/},
  \href{http://dx.doi.org/10.1140/epjc/s10052-015-3363-3}{Eur.Phys.J. {\bf C75}
  (2015) no.~5, 228},
\href{http://arxiv.org/abs/1411.1855}{{\tt arXiv:1411.1855 [hep-ex]}}.

\bibitem{Aad:2014vwa}
{ATLAS} Collaboration, G.~Aad et al., {\em {Measurement of the inclusive jet
  cross-section in proton-proton collisions at $\sqrt{s}=7$ TeV using
  4.5~fb$^{-1}$ of data with the ATLAS detector}\/},
\href{http://arxiv.org/abs/1410.8857}{{\tt arXiv:1410.8857 [hep-ex]}}.

\bibitem{Aad:2014pua}
{ATLAS} Collaboration, G.~Aad et al., {\em {Measurements of jet vetoes and
  azimuthal decorrelations in dijet events produced in $pp$ collisions at
  $\sqrt{s}$ = 7 TeV using the ATLAS detector}\/},
  \href{http://dx.doi.org/10.1140/epjc/s10052-014-3117-7}{Eur.Phys.J. {\bf C74}
  (2014) no.~11, 3117},
\href{http://arxiv.org/abs/1407.5756}{{\tt arXiv:1407.5756 [hep-ex]}}.

\bibitem{Aad:2013tea}
{ATLAS} Collaboration, G.~Aad et al., {\em {Measurement of dijet cross sections
  in $pp$ collisions at 7 TeV centre-of-mass energy using the ATLAS
  detector}\/},  \href{http://dx.doi.org/10.1007/JHEP05(2014)059}{JHEP {\bf
  1405} (2014)  059},
\href{http://arxiv.org/abs/1312.3524}{{\tt arXiv:1312.3524 [hep-ex]}}.

\bibitem{Aad:2012dr}
{ATLAS} Collaboration, G.~Aad et al., {\em {Measurements of the pseudorapidity
  dependence of the total transverse energy in proton-proton collisions at
  $\sqrt{s} = 7$ TeV with ATLAS}\/},
\href{http://arxiv.org/abs/1208.6256}{{\tt arXiv:1208.6256 [hep-ex]}}.

\bibitem{Aad:2012meb}
{ATLAS} Collaboration, G.~Aad et al., {\em {ATLAS measurements of the
  properties of jets for boosted particle searches}\/},
  \href{http://dx.doi.org/10.1103/PhysRevD.86.072006}{Phys.Rev. {\bf D86}
  (2012)  072006},
\href{http://arxiv.org/abs/1206.5369}{{\tt arXiv:1206.5369 [hep-ex]}}.

\bibitem{Aad:2011fc}
{ATLAS} Collaboration, G.~Aad et al., {\em {Measurement of inclusive jet and
  dijet production in $pp$ collisions at $\sqrt{s} = 7$ TeV using the ATLAS
  detector}\/},
\href{http://arxiv.org/abs/1112.6297}{{\tt arXiv:1112.6297 [hep-ex]}}.

\bibitem{Collaboration:2011tq}
{ATLAS} Collaboration, G.~Aad et al., {\em {Measurement of multi-jet cross
  sections in proton-proton collisions at a 7 TeV center-of-mass energy}\/},
\href{http://arxiv.org/abs/1107.2092}{{\tt arXiv:1107.2092 [hep-ex]}}.

\bibitem{Aad:2011jz}
{ATLAS} Collaboration, G.~Aad et al., {\em {Measurement of dijet production
  with a veto on additional central jet activity in $pp$ collisions at
  $\sqrt{s} = 7$ TeV using the ATLAS detector}\/},
  \href{http://arxiv.org/abs/1107.1641}{{\tt arXiv:1107.1641 [hep-ex]}}.

\bibitem{Aad:2011ni}
{ATLAS} Collaboration, G.~Aad et al., {\em {Measurement of dijet azimuthal
  decorrelations in $pp$ Collisions at $\sqrt{s} = 7$ TeV}\/},
\href{http://arxiv.org/abs/1102.2696}{{\tt arXiv:1102.2696 [hep-ex]}}.

\bibitem{Aad:2011kq}
{ATLAS} Collaboration, G.~Aad et al., {\em {Study of jet shapes in inclusive
  jet production in $pp$ collisions at $\sqrt{s}=7$~TeV using the ATLAS
  detector}\/},  \href{http://dx.doi.org/10.1103/PhysRevD.83.052003}{Phys. Rev.
  {\bf D83} (2011)  052003},
\href{http://arxiv.org/abs/1101.0070}{{\tt arXiv:1101.0070 [hep-ex]}}.

\bibitem{Aad:2010wv}
{ATLAS} Collaboration, G.~Aad et al., {\em {Measurement of inclusive jet and
  dijet cross sections in proton-proton collisions at 7 TeV centre-of-mass
  energy with the ATLAS detector}\/},
\href{http://arxiv.org/abs/1009.5908}{{\tt arXiv:1009.5908 [hep-ex]}}.

\bibitem{Chatrchyan:2014gia}
{CMS} Collaboration, S.~Chatrchyan et al., {\em {Measurement of the ratio of
  inclusive jet cross sections using the anti-$k_T$ algorithm with radius
  parameters $R=0.5$ and $0.7$ in $pp$ collisions at $\sqrt{s}$ = 7 TeV}\/},
\href{http://arxiv.org/abs/1406.0324}{{\tt arXiv:1406.0324 [hep-ex]}}.

\bibitem{Chatrchyan:2013rla}
{CMS} Collaboration, S.~Chatrchyan et al., {\em {Studies of jet mass in dijet
  and $W$/$Z$+jet events}\/},
\href{http://arxiv.org/abs/1303.4811}{{\tt arXiv:1303.4811 [hep-ex]}}.

\bibitem{Chatrchyan:2012bja}
{CMS} Collaboration, S.~Chatrchyan et al., {\em {Measurements of differential
  jet cross sections in proton-proton collisions at $\sqrt{s}=7$ TeV with the
  CMS detector}\/},
  \href{http://dx.doi.org/10.1103/PhysRevD.87.112002}{Phys.Rev. {\bf D87}
  (2013) no.~11, 112002},
\href{http://arxiv.org/abs/1212.6660}{{\tt arXiv:1212.6660 [hep-ex]}}.

\bibitem{Chatrchyan:2012vc}
{CMS} Collaboration, S.~Chatrchyan et al., {\em {Observation of a diffractive
  contribution to dijet production in proton-proton collisions at $\sqrt{s}=7$
  TeV}\/},  \href{http://dx.doi.org/10.1103/PhysRevD.87.012006}{Phys. Rev. {\bf
  D87} (2013) no.~1, 012006},
\href{http://arxiv.org/abs/1209.1805}{{\tt arXiv:1209.1805 [hep-ex]}}.

\bibitem{Chatrchyan:2012bf}
{CMS} Collaboration, S.~Chatrchyan et al., {\em {Search for quark compositeness
  in dijet angular distributions from $pp$ collisions at $\sqrt{s}=7$ TeV}\/},
  \href{http://dx.doi.org/10.1007/JHEP05(2012)055}{JHEP {\bf 1205} (2012)
  055},
\href{http://arxiv.org/abs/1202.5535}{{\tt arXiv:1202.5535 [hep-ex]}}.

\bibitem{Chatrchyan:2012gwa}
{CMS} Collaboration, S.~Chatrchyan et al., {\em {Measurement of the inclusive
  production cross sections for forward jets and for dijet events with one
  forward and one central jet in $pp$ collisions at $\sqrt{s} = 7$ TeV}\/},
  \href{http://dx.doi.org/10.1007/JHEP06(2012)036}{JHEP {\bf 1206} (2012)
  036},
\href{http://arxiv.org/abs/1202.0704}{{\tt arXiv:1202.0704 [hep-ex]}}.

\bibitem{Chatrchyan:2011wm}
{CMS} Collaboration, S.~Chatrchyan et al., {\em {Measurement of energy flow at
  large pseudorapidities in $pp$ collisions at $\sqrt{s}$ = 0.9 and 7 TeV}\/},
  \href{http://dx.doi.org/10.1007/JHEP11(2011)148}{JHEP {\bf 11} (2011)  148},
\href{http://arxiv.org/abs/1110.0211}{{\tt arXiv:1110.0211 [hep-ex]}}.

\bibitem{Chatrchyan:2011wn}
{CMS} Collaboration, S.~Chatrchyan et al., {\em {Measurement of the ratio of
  the 3-jet to 2-jet cross sections in $pp$ collisions at $\sqrt{s} = 7$
  TeV}\/},  Phys.Lett. {\bf B702} (2011)  336--354,
  \href{http://arxiv.org/abs/1106.0647}{{\tt arXiv:1106.0647 [hep-ex]}}.

\bibitem{Chatrchyan:2011me}
{CMS} Collaboration, S.~Chatrchyan et al., {\em {Measurement of the inclusive
  jet cross section in $pp$ collisions at $\sqrt{s} = 7$ TeV}\/},
  \href{http://dx.doi.org/10.1103/PhysRevLett.107.132001}{Phys. Rev. Lett. {\bf
  107} (2011)  132001}, \href{http://arxiv.org/abs/1106.0208}{{\tt
  arXiv:1106.0208 [hep-ex]}}. Long author list - awaiting processing.

\bibitem{Khachatryan:2011as}
{CMS} Collaboration, V.~Khachatryan et al., {\em {Measurement of dijet angular
  distributions and search for quark compositeness in $pp$ collisions at
  7~\TeV}\/},  \href{http://dx.doi.org/10.1103/PhysRevLett.106.201804}{Phys.
  Rev. Lett. {\bf 106} (2011)  201804},
\href{http://arxiv.org/abs/1102.2020}{{\tt arXiv:1102.2020 [hep-ex]}}.

\bibitem{Khachatryan:2011zj}
{CMS} Collaboration, V.~Khachatryan et al., {\em {Dijet azimuthal
  decorrelations in $pp$ collisions at $\sqrt{s} = 7$~TeV}\/},
  \href{http://dx.doi.org/10.1103/PhysRevLett.106.122003}{Phys.Rev.Lett. {\bf
  106} (2011)  122003}, \href{http://arxiv.org/abs/1101.5029}{{\tt
  arXiv:1101.5029 [hep-ex]}}.

\bibitem{Catani:1990rr}
S.~Catani, B.~Webber, and G.~Marchesini, {\em {QCD coherent branching and
  semi-inclusive processes at large $x$}\/},
\href{http://dx.doi.org/10.1016/0550-3213(91)90390-J}{Nucl.Phys. {\bf B349}
  (1991)  635--654}.

\bibitem{Buckley:2009bj}
A.~Buckley, H.~Hoeth, H.~Lacker, H.~Schulz, and J.~E. von Seggern, {\em
  {Systematic event generator tuning for the LHC}\/},
  \href{http://dx.doi.org/10.1140/epjc/s10052-009-1196-7}{Eur. Phys. J. {\bf
  C65} (2010)  331--357},
\href{http://arxiv.org/abs/0907.2973}{{\tt arXiv:0907.2973 [hep-ph]}}.

\bibitem{ATL-PHYS-PUB-2014-021}
{ATLAS Collaboration}, {\em {ATLAS Run~1 \PythiaEight tunes}\/},  Tech. Rep.
  ATL-PHYS-PUB-2014-021, CERN, Geneva, Nov, 2014.
\newblock \url{https://cds.cern.ch/record/1966419}.

\bibitem{Skands:2014pea}
P.~Skands, S.~Carrazza, and J.~Rojo, {\em {Tuning PYTHIA 8.1: the Monash 2013
  Tune}\/},  \href{http://dx.doi.org/10.1140/epjc/s10052-014-3024-y}{Eur. Phys.
  J. {\bf C74} (2014) no.~8, 3024},
\href{http://arxiv.org/abs/1404.5630}{{\tt arXiv:1404.5630 [hep-ph]}}.

\end{thebibliography}\endgroup

\end{document}